\documentclass[journal,10pt]{IEEEtran}

\usepackage[utf8]{inputenc}
\usepackage[T1]{fontenc}
\usepackage{textcomp}
\usepackage{fixltx2e}
\usepackage{comment}

\usepackage[pdftex]{graphicx}
\DeclareGraphicsExtensions{.pdf,.png,.jpg}
\usepackage{soul}
\usepackage[caption=false,font=footnotesize]{subfig}
\usepackage{booktabs}
\usepackage{url}
\usepackage{layouts}
\usepackage{amsmath}
\usepackage{amsthm}
\usepackage{amssymb}
\usepackage{acronym}
\usepackage[capitalise,noabbrev]{cleveref}
\usepackage[textwidth=1cm]{todonotes}
\usepackage{xspace}
\usepackage{booktabs}
\usepackage{algorithm,algorithmic}
\usepackage{graphicx}
\usepackage{csquotes}

\usepackage[american]{babel}
\usepackage[load-configurations=binary,detect-all,binary-units=true,range-phrase=--,per-mode=symbol]{siunitx}
\DeclareSIUnit\bit{bit}
\DeclareSIUnit\byte{Byte}
\DeclareSIUnit\decibeli{dBi}
\DeclareSIUnit\decibelm{dBm}

\acrodef{ACC}{Adaptive Cruise Control}
\acrodef{AFS}{Adaptive Front-Lighting Systems}
\acrodef{ARIB}{Association of Radio Industries and Businesses}
\acrodef{ATB}{Adaptive Traffic Beacon}
\acrodef{CACC}{Cooperative Adaptive Cruise Control}
\acrodef{CAM}{Cooperative Awareness Message}
\acrodef{CBR}{Channel Busy Ratio}
\acrodef{CPS}{Cyber-Physical System}
\acrodef{CPSS}{Cyber-Physical-Social System}
\acrodef{C-V2X}{Cellular V2X}
\acrodef{D2D}{Device-to-Device}
\acrodef{DCC}{Decentralized Congestion Control}
\acrodef{DEB}{Distributed EDCA Bursting}
\acrodef{DSRC}{Dedicated Short-Range Communication}
\acrodef{ETSI}{European Telecommunications Standards Institute}
\acrodef{FDR}{Full-Duplex Relaying}
\acrodef{FSM}{Finite State Machine}
\acrodef{H-CPS}{Hybridized Cyber-Physical System}
\acrodef{ITS}{Intelligent Transportation System}
\acrodef{LQR}{Linear Quadratic Regulator}
\acrodef{MIMO}{Multiple-Input Multiple-Output}
\acrodef{mmWave}{Millimeter Wave}
\acrodef{MPC}{Model Predictive Control}
\acrodef{RADCOM}{Radar-based Communication}
\acrodef{RF}{Radio Frequency}
\acrodef{TRC}{Transmit Rate Control}
\acrodef{V2V}{Vehicle-to-Vehicle}
\acrodef{V2X}{Vehicle-to-Everything}
\acrodef{VLC}{Visible Light Communication}

\usepackage[color]{changebar}
\def\todoCtd#1{%
	TODO: #1%
	\ifx&#1&...\fi%
	\endgroup
	\cbend
	\relax
}

\NewDocumentCommand\IEEE{ s m d[] }{%
	\IfBooleanTF{#1}{}{IEEE\,}
	\nolinebreak[2]
	#2%
	\IfNoValueTF{#3}{%
	}{%
		\StrGobbleLeft{#3}{1}[\sommerIEEEFirstLetter]%
		\IfEq{\sommerIEEEFirstLetter}{}{%
			#3
		}{%
			\nolinebreak[3]
			\StrLeft{#3}{1}%
			\sommerIEEELettersSlashed{\sommerIEEEFirstLetter}%
		}%
	}%
}
\newcommand{\sommerIEEELettersSlashed}[1]{%
	/
	\StrLeft{#1}{1}%
	\StrGobbleLeft{#1}{1}[\sommerIEEESubsequentLetter]%
	\IfEq{\sommerIEEESubsequentLetter}{}{%
	}{%
		\sommerIEEELettersSlashed{\sommerIEEESubsequentLetter}
	}%
}

\newcommand{\p}{\IEEE{802.11}[p]}

\begin{document}
\title{Centralized Model Predictive Control with Human-Driver Interaction for Platooning}

\author{%
	Justin M. Kennedy~\IEEEmembership{Member, IEEE},
	Julian Heinovski~\IEEEmembership{Student Member, IEEE},
	\\
	Daniel E.\ Quevedo~\IEEEmembership{Fellow, IEEE},
	and
	Falko Dressler~\IEEEmembership{Fellow, IEEE}
\thanks{Copyright (c) 2023 IEEE. Personal use of this material is permitted. However, permission to use this material for any other purposes must be obtained from the IEEE by sending a request to pubs-permissions@ieee.org.}
\thanks{%
This work has been supported in part by the project NICCI2 funded by the German Research Foundation (DFG) under grant numbers DR 639/23-2 and QU 437/1-2.
}
\thanks{%
Justin M. Kennedy and Daniel E. Quevedo are with the School of Electrical Engineering and Robotics, Queensland University of Technology, Australia;
Julian Heinovski and Falko Dressler are with the School of Electrical Engineering and Computer Science, TU Berlin, Germany;
(\{j12.kennedy, daniel.quevedo\}@qut.edu.au,
\{heinovski, dressler\}@ccs-labs.org).
}}%

\maketitle

\begin{abstract}

Cooperative adaptive cruise control presents an opportunity to improve road transportation through increase in road capacity and reduction in energy use and accidents.
Clever design of control algorithms and communication systems is required to ensure that the vehicle platoon is stable and meets desired safety requirements.
In this paper, we propose a centralized model predictive controller for a heterogeneous platoon of vehicles to reach a desired platoon velocity and individual inter-vehicle distances with driver-selected headway time.
As a novel concept, we allow for interruption from a human driver in the platoon that temporarily takes control of their vehicle with the assumption that the driver will, at minimum, obey legal velocity limits and the physical performance constraints of their vehicle.
The finite horizon cost function of our proposed platoon controller is inspired from the infinite horizon design.
To the best of our knowledge, this is the first platoon controller that integrates human-driven vehicles.
We illustrate the performance of our proposed design with a numerical study, demonstrating that the safety distance, velocity, and actuation constraints are obeyed.
Additionally, in simulation we illustrate a key property of string stability where the impact of a disturbance is reduced through the platoon.

\end{abstract}


\acresetall
\IEEEpeerreviewmaketitle

%

\section{Introduction}%
\label{sec:intro}

Autonomous vehicle platooning with inter-vehicle communication permits road vehicles to travel close together increasing road capacity while reducing energy use and associated vehicle emissions \cite{sommer2014vehicular}.
This cooperative connected cruise control technology can reduce the incidence of so-called ghost traffic jams \cite{Wang2020SurveyCooperativeLongitudinal} and highway accidents \cite{Kunze2010OrganizationOperationElectronically}.
The autonomous cruise control problem was first posed as a centralized platoon design approach in \cite{Levine1966IEEETransactionsonAutomaticControloptimalerrorregulation}, and has seen recent attention with several survey papers \cite{Alam2015IEEEControlSystemsHeavyDutyVehicle,Wang2020SurveyCooperativeLongitudinal,Besselink2016CyberPhysicalControlRoad}.

\begin{figure}
    \centering
    \includegraphics[width=\columnwidth]{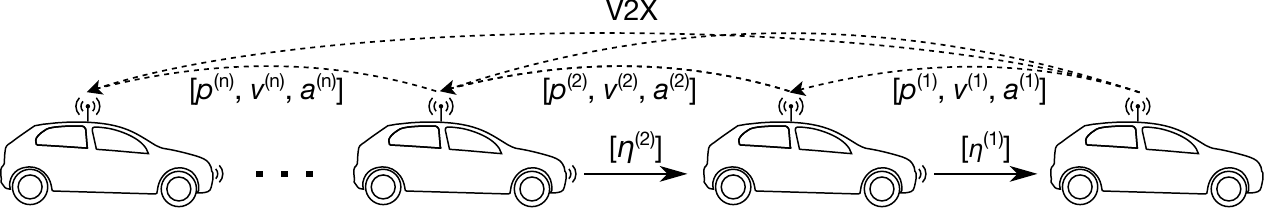}
    \caption{System model of the used \acs{CACC} concept: a platoon controller regulates the velocity and inter-vehicle distances utilizing the vehicles' position $p$, velocity $v$, and acceleration $a$, exchanged via inter-vehicle communications.}
    \label{fig:cacc-concept}
\end{figure}

\ac{ACC} systems use on board sensors to measure the distance and velocity of a predecessor vehicle to operate an autonomous cruise control system.
However, these systems are prone to string stability issues resulting in ghost traffic jams \cite{Swaroop1996Stringstabilityinterconnected}.
To ensure stability of the platoon, either a large, velocity-based inter-vehicle distance is required using a headway time \cite{Wu2020IEEEAccessSpacingPoliciesAdaptive} or more than just the preceding vehicle's state is required.
By utilizing inter-vehicle communication systems, \ac{CACC} systems can reduce the inter-vehicle distance and avoid string stability issues by sharing the desired control actions from other vehicles.
Figure~\ref{fig:cacc-concept} illustrates our system model for communication, which represents a potential \ac{CACC} design of a coordinated platoon of vehicles driving with small inter-vehicle distances with a platoon communication network.

Many platoon control designs have focused on the technical aspects of algorithm design, communication constraints, or experimentation isolated from real traffic \cite{Lank2011InteractionHumanMachine}.
However, studies of driver behavior have found that human drivers maintain small inter-vehicle distances below safety margins \cite{Martinez2007IEEETransactionsonControlSystemsTechnologySafeLongitudinalControl}.
This effect is exacerbated when humans drive when near autonomous platoons \cite{Gouy2014Drivingnextautomated}.
As pointed out in \cite{Dey2016IEEETransactionsonIntelligentTransportationSystemsReviewCommunicationDriver}, during implementation of vehicle platooning systems, it will be important to incorporate legacy vehicles that are unable to integrate with a platoon communication system and human-driver interaction for passenger comfort and well-being, e.g., some passengers feel uncomfortable with too small safety gaps, as well as motion sickness.
Additionally, some drivers may wish to switch automation levels, connecting or disconnecting from the \ac{CACC}.
The inclusion of human factors is in line with research integrating human behavior and \ac{CPS}, moving from \ac{CPS} to \ac{H-CPS} which is also called \ac{CPSS} \cite{dressler2018cpss}.

While modern communication systems can enable the transmission of large volumes of data very quickly between physically separated vehicles \cite{amjad2020inband}, the inherent reliability limitations of wireless communications systems, including from packet delays and dropouts, can degrade the information available to a control system.
When implementing a \ac{CACC} system for a practical vehicle platoon, the information available to the local control system at each individual vehicle should be considered.
When ignoring communication imperfections such as delays and packet loss, a centralized platoon control design that incorporates constraints and guarantees safe inter-vehicle distances could lead to optimal decisions.
Therefore, dealing with realistic wireless communications, a number of decentralized and distributed control designs have been proposed using local information in combination with information transmitted from neighboring cars (usually those in front), cf.\ \cite{Wang2020SurveyCooperativeLongitudinal,Alam2015IEEEControlSystemsHeavyDutyVehicle,Wu2020IEEEAccessSpacingPoliciesAdaptive,Dey2016IEEETransactionsonIntelligentTransportationSystemsReviewCommunicationDriver}.
However, the lack of full information of all platoon members means that the optimality of the control actions may be degraded.
This motivates the development of decentralized control architectures.
One aim in the design of decentralized controllers is to recover as close as possible the performance provided by a centralized controller that has access to all system and state information \cite{Farina2019DistributedMPCLarge}.
While many decentralized \ac{CACC} designs have been posed, a centralized platoon control design that incorporates constraints and guarantees safe inter-vehicle distances has not yet been presented.
This is of particular interest, when dealing with heterogeneity and human-driver interaction, which cannot easily be realized with incomplete information in a distributed way.
Thus, to enable comparison to decentralized or distributed control designs with a limited or reduced information availability, it is first necessary to understand centralized \ac{CACC} designs.

\begin{figure}
    \centering
    \includegraphics[width=\columnwidth]{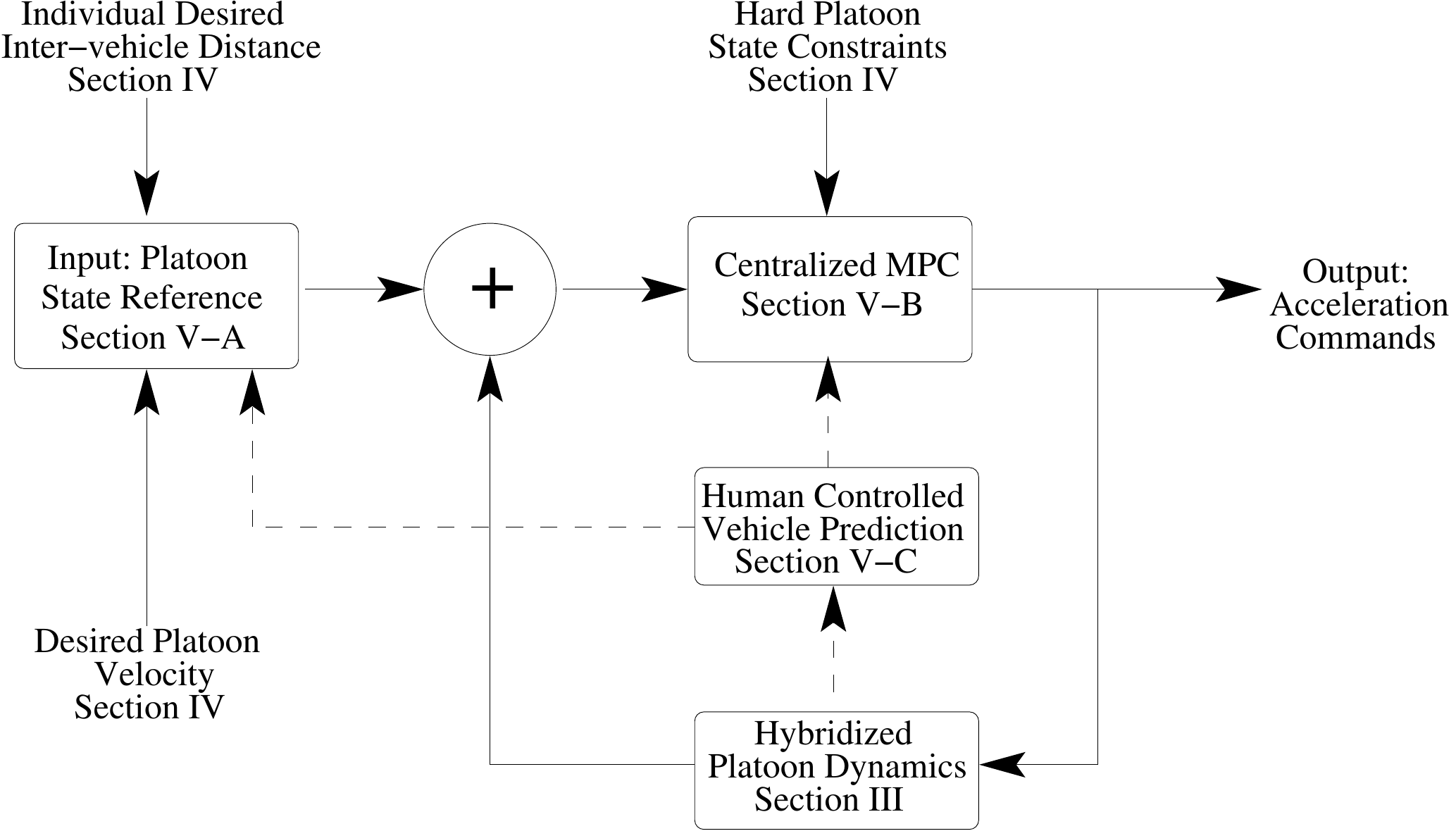}
    \caption{Platoon control architecture. The exogenous inputs to the controller are the desired inter-vehicle distances and platoon velocity to form the platoon reference. The output is the control action for all vehicles in the platoon. The controller reconfigures when a human-driver temporarily takes control. The block connected by the dashed lines engages predicting a minimum control action of the human-driver.}
    \label{fig:controlloop}
\end{figure}

In this paper, we present a novel constrained \ac{MPC} approach for the centralized control of a platoon of heterogeneous vehicles.
In our design, we include reconfiguration under temporary human-driver control.
The goal of our controller is to reach a desired platoon velocity and individual inter-vehicle distances, while guaranteeing safety constraints.
The desired inter-vehicle distance is based on driver-selected headway times which are variable between individual vehicles in the platoon, and can change over time \cite{dressler2019cooperative,milanes2014cooperative,segata2015communication}.
Figure~\ref{fig:controlloop} illustrates the control loop of our proposed centralized platoon controller.
The exogenous inputs to our controller are the desired platoon velocity and individual inter-vehicle distances, and the desired safety constraints on the inter-vehicle distance, velocity, and acceleration.
The output from our controller is the desired control action for every vehicle in the platoon.

In \ac{MPC} design, the control law is based on a prediction of the states of the system.
To compute a mathematical prediction of the system states, we utilize an abstract mathematical model of the system dynamics.
This technique is widely used in control engineering practice \cite{Badgwell2015ModelPredictiveControl} and \cite{Qin2003ControlEngineeringPracticesurveyindustrialmodel}.

To ensure convergence of the vehicle positions and velocities to the desired references over a finite horizon, we design a time-varying reference for all vehicles in the platoon.
We consider a combined mathematical model of the platoon of individual vehicle dynamics to predict the motion of all of the vehicles from the multiple control inputs.
Inspired by the infinite horizon optimal control algorithm \cite{Jovanovic2005IEEETransactionsonAutomaticControlillposednesscertain}, we propose a finite horizon cost function.
We introduce time-varying references to encode the desired trajectories of positions, velocities, and accelerations for all vehicles in the platoon.
We emphasize that in our design, the desired inter-vehicle distance and platoon velocity are allowed to change over time.
This feature enables a controlled dynamic behavior of the platoon.
Our cost function penalizes the quadratic error of the predicted vehicle positions, velocities, and accelerations to the desired time-varying references, as well as the predicted inter-vehicle distances to the desired inter-vehicle distances.
The use of a quadratic cost function allows for simple application of hard constraints on the cost function in a quadratic program optimizer to guarantee minimum and maximum inter-vehicle distances, velocities, and accelerations of all vehicles in the platoon.
The inclusion of constraints in the controller ensures that the control actions do not result in unsafe vehicle behavior, including reversing, extreme accelerations, and most critically, collisions between multiple vehicles.

In our design, we include a switch to reconfigure our controller for a vehicle temporarily under human driver control, shown as the dashed line in Figure~\ref{fig:controlloop}.
Early implementations of potential \ac{CACC} designs will need to operate with non-\acs{CACC} enabled vehicles and with drivers that choose to switch automation levels \cite{Dey2016IEEETransactionsonIntelligentTransportationSystemsReviewCommunicationDriver}.
Our approach accommodates a human driver to make an emergency brake, reduce speed, or travel at a different inter-vehicle distance.
Our method could also be utilized to incorporate legacy vehicles.
When a vehicle is human driver controlled, we adjust our time-varying reference to return the platoon to the desired reference and remove the vehicle from the platoon control.
A key feature of our method is that it incorporates the vehicle states such that the control actions for the platoon still ensure the inter-vehicle distance constraints for the other vehicles in the platoon.
We predict the control actions of the human driver by assuming that the driver will only change their control action to obey a minimum set of constraints of the legal road speed limits and their vehicle performance constraints.
Our centralized platoon control design ensures safety for the remaining platoon-controlled vehicles by producing control actions that maintain safe inter-vehicle distances.
More complex predictive approaches, such as economic cost functions \cite{Fleming2021IncorporatingDriverPreferences} or inter-vehicle interactions, could be utilized to predict a human driver.

The key contributions of this paper are
\begin{itemize}
	\item proposal of an optimal algorithm for centralized platoon control of heterogeneous vehicles with constraints, and
	\item reconfiguration under temporary human-driver control by incorporation of a human-driver model.
\end{itemize}

%

\section{Related Work}%
\label{sec:related}

\acused{ETSI}
\acused{DRSC}
\acused{ARIB}

In this paper, we propose an \ac{MPC} design to perform the \ac{CACC} task to safely control a platoon of vehicles to the desired inter-vehicle distances and velocity.
Through this section, we present the literature on communication protocols and control law designs for \ac{CACC}, to argue that \ac{MPC} is an appropriate technology for \ac{CACC} with interruption by human drivers.
We highlight that while many control designs have been posed, a centralized control design that obeys safety constraints and incorporates human drivers has yet to be posed.

\subsection{CACC Communication}

In order to work reliably and with small safety gaps, \ac{CACC} requires periodic updates of vehicles' data (e.g., acceleration, speed, position).
Typically, the data from at least the vehicle in front and often also the first vehicle (i.e., the platoon leader) is necessary.
If the updates arrive with a high enough frequency for the control system to react properly, string-stability, i.e., keeping the desired gaps without accumulating control errors throughout platoon members, can be achieved \cite{fernandes2012platooning}.
If the updates are delayed, {``string-stability is seriously compromised''}~\cite{liu2001effects}.

\subsubsection{\p{}}
Up to a few years ago, the main \ac{V2X} technology considered was \p{} as a basis for quite advanced protocol families such as \ac{ETSI} ITS-G5 \cite{sommer2014vehicular}.

The most simple approach for exchanging the vehicle updates is to use \emph{static beaconing}, where vehicles broadcast their information in regular, periodic intervals.
Yet, static beaconing can lead to a congested channel, especially in highly dense scenarios, e.g., with long or many platoons, thus reducing the stability of a platoon.
Thus, Segata et al. \cite{segata2015communication}
proposed to use \emph{slotted beaconing}, which splits the time for the leader beacon into transmission slots for all platoons members.
The authors show that this can greatly improve the beaconing performance in crowded scenarios, especially when combined with transmit power control, thus, reducing the load and improving the reliability.

In order to reduce the channel load further, \emph{dynamic beaconing} schemes have been proposed.
Sommer et al. \cite{sommer2011traffic} presented the \ac{ATB} protocol, which adaptively adjusts the beaconing period according to the current channel quality and the message utility.
Following up on this, \ac{ETSI} standardized \ac{DCC}~\cite{etsi_ts_102687_v111}.
It uses a simple final state machine to adjust, among others, beacon interval and transmit power based on the current observed channel busy ratio.
Sommer et al. \cite{sommer2015how} proposed \emph{DynB} to avoid overloading the wireless channel and allow low-latency communication by using very short beaconing intervals.
The protocol continuously observes the channel load and considers detailed radio shadowing effects, even by moving vehicles, that block the transmission and the number of neighboring vehicles to calculate the best beacon interval.
Focusing specifically on platooning, Segata et al. \cite{segata2015jerk} proposed a dynamic approach called \emph{jerk beaconing}, which exploits vehicle dynamics to share data only when needed by the controller.
Here, the beacon interval is computed dynamically based on changes in acceleration over time, i.e., the jerk.
This approach shows huge benefits in terms of network resource saving and is able to keep inter-vehicle distance close to the desired gap even in highly demanding scenarios.

 Going beyond \p{}, Segata et al. \cite{segata2017lets} proposed the \ac{DEB} protocol, which extends the frame busting mechanism of \p{} such that only the platoon leaders content for the channel.
In case of successful channel reservation, all vehicles in the platoon transmit a coordinated burst of frames, thus, sharing the platoon leader's transmission opportunity.
This helps overcoming channel limits by reducing the number of nodes contending for the channel and improving spatial reuse.
Amjad et al. \cite{amjad2019platooning} extend \p{} by adding a full-duplex relaying system, which enables platoon members to simultaneously receive and relay the leader beacons.

\subsubsection{\acf{C-V2X}}%

Albeit all of the above protocols and modifications, \p{} alone seems not to be sufficient for meeting the strict requirements of \ac{CACC} (i.e., ultra-low reliability and latency) \cite{segata2022multi-technology}.
The most prominent alternative for enabling \ac{V2X} communication is \ac{C-V2X}, which uses 3GPP standardized 5G cellular networks.
Radio resources are scheduled by either the base station if vehicles are in-coverage (operation mode 3) or by a distributed resource allocation scheme if vehicles are out-of-coverage (operation mode 4).
The latter allows vehicles to select resources in a stand-alone fashion with \emph{semi-persistent scheduling}.
While mode 3 in general allows for high packet receptions ratios, mode 4 produces lower beacon update delays \cite{cecchini2017performance}, which are also required for platooning.

For example, Vukadinovic et al. \cite{vukadinovic2018three} compare \p{} to 3GPP \ac{C-V2X} based on LTE in both operation modes for truck platooning.
Results show that \ac{C-V2X} in both modes allows for shorter inter-truck distances than \p{} due to more reliable communication in a congested wireless channel.
However, short communication distances and large vehicle densities seem to be covered better with \p{} instead of \ac{C-V2X} \cite{bazzi2017performance}.
Therefore, general modifications for improving the scheduling of sidlelink radio resources in mode 4 have been proposed \cite{molina2017lte,gonzalez2019feasibility}.
In order to reach the performance required for \ac{CACC}, Hegde et al. \cite{hegde2019enhanced} propose to schedule the sidelink radio resources for the platoon members by the platoon leader.
Similarly, the radio resource coordination method by Campolo et al. \cite{campolo2017better} fulfills the ultra-low latency requirements of \ac{CACC} and is able to provide spatial reuse of LTE resources among platoon members.

\subsubsection{RADCOM}

Complementary to \p{} and \ac{C-V2X}, joint communication and sensing approaches, also known as \ac{RADCOM}, have been proposed.
Following the trend of using higher communication frequencies for radio communication, \ac{mmWave} technologies have recently become interesting to the \ac{V2X} research community.
\ac{mmWave} technology promises high bitrates and low delays due to its wide channel bandwidth and dynamic beam-forming \cite{giordani2017millimeter}.
However, using it as a single communication technology may be difficult due to its highly volatile transmission channel, especially in an automotive environment \cite{giordani2017millimeter,dimce2021mmwave}.
Nevertheless, initial works indicate that \ac{mmWave} can be very valuable when complementing the other alternatives \cite{coll-perales2019sub-6ghz,amjad2020inband}.

%

\subsection{CACC Controller}

Control design for vehicle platooning has focused on meeting string stability conditions with several definitions in the literature \cite{Feng2019AnnualReviewsinControlStringstabilityvehicular}.
In addition to stability requirements, the control design also needs to consider the information flow topology arising from the available communication links, formation geometry or spacing policy, vehicle dynamics, and desired platoon convergence.

The information flow topology of how information is shared between vehicles influences both the control algorithm design and the required communication system.
Many control designs utilize a leader-follower approach where a lead vehicle sets the platoon speed and each follower vehicle maintains their own spacing to the predecessor, such as the sliding mode controller in \cite{rajamani2012vehicle} and employed in \cite{segata2015communication}.
Other designs consider bi-directional information sharing from the neighboring vehicles such that leader information is not required, e.g. \cite{Knorn2014AutomaticaPassivitybasedcontrol} and \cite{Hu2020CooperativeControlHeterogeneous}.
These distributed approaches consider that the lead or reference vehicle is exogenous to the platoon controller \cite{Zhou2012VehiclePlatoonControl}, either controlled by a human driver or by a separate \ac{ACC} system \cite{segata2015communication}.
Many designs focus on \ac{V2V}, also of interest are \ac{V2X} where infrastructure can monitor and coordinate a platoon \cite{Alam2015IEEEControlSystemsHeavyDutyVehicle}, as well as interactions with other platoons \cite{Wang2018Infrastructureassistedadaptive}.
Recent works have included unreliable communication channels in the control design using time delays \cite{santini2017consensusbased} and packet loss \cite{giordano2019joint}.

A key design factor is the formation geometry of the inter-vehicle distances \cite{Wu2020IEEEAccessSpacingPoliciesAdaptive}, which was considered fixed in early works \cite{Levine1966IEEETransactionsonAutomaticControloptimalerrorregulation} but caused string instability for \ac{ACC} \cite{Feng2019AnnualReviewsinControlStringstabilityvehicular}.
To achieve string stability for \ac{ACC}, \cite{Chien1992AutomaticVehicleFollowing} proposed a velocity based spacing policy following the concept that a human driver should follow a preceding car with a certain headway time, with refinements in \cite{Ioannou1993IEEETransactionsonVehicularTechnologyAutonomousintelligentcruise} and \cite{SWAROOP1994VehicleSystemDynamicsComparisionSpacingHeadway}.
To account for the slower braking performance of heavy vehicles, a variable spacing policy with the headway a function of the difference in velocity \cite{Yanakiev1998IEEETransactionsonVehicularTechnologyNonlinearspacingpolicies}.
The authors note that if lead vehicle information is shared, then the headway is able to be reduced to zero \cite{Yanakiev1995Longitudinalcontrolheavy}.
Often a common (non-unique) and constant headway time is utilized \cite{ploeg2011design}.
Alternative spacing policies have also included use of the traffic density \cite{Tak2016IEEETransactionsonIntelligentTransportationSystemsStudyTrafficPredictive},
as constant time headway can result in unstable traffic flow  \cite{Wang2004IEEETransactionsonVehicularTechnologyShouldAdaptiveCruise}.

The vehicle dynamics used for control design of vehicle platoons have included complex models that model torque output of the engine with variable gear ratios as well as simplified linear models.
In reference to the nonlinear engine and gearbox models, it is noted in \cite{rajamani2012vehicle} that a first order lag model is suitable for higher level control of the vehicle, such as for platooning applications.
This simplifies the vehicle, engine, and braking systems into a single constant.
In \cite{rajamani2012vehicle,segata2015communication} the mechanical lag coefficient for a standard passenger vehicle is assumed to be $\tau_i = 0.5$ [s], with a heavy vehicle having a larger coefficient.
Alternative modeling approaches have used the energy based port-Hamiltonian system model \cite{Knorn2014AutomaticaPassivitybasedcontrol,Silva2021AutomaticaStringstableintegral}.
While use of a homogeneous platoon with identical dynamics makes the control design and tuning simpler, it is unrealistic to real world heterogeneous platoons of different vehicles \cite{Gao2016Robustcontrolheterogeneous}.
Certain controller stability properties can change with different types of vehicles such as platoons of heavy vehicles \cite{Alam2014Guaranteeingsafetyheavy}, and environment effects including changes in road slope \cite{Turri2017CooperativeLookAhead} and wind \cite{Gao2016Robustcontrolheterogeneous}.

A variety of control design approaches have been proposed in the literature including the classic \ac{LQR} \cite{Levine1966IEEETransactionsonAutomaticControloptimalerrorregulation,Hu2020CooperativeControlHeterogeneous,Alam2015Experimentalevaluationdecentralized}, Proportional Integral Derivative \cite{Fiengo2019DistributedRobustPID}, H-infinity \cite{Gao2016Robustcontrolheterogeneous}, sliding mode control \cite{rajamani2012vehicle}, and \ac{MPC}.
\ac{MPC} algorithms optimize a finite horizon cost function at each time step, and allow for the inclusion of hard constraints \cite{Maciejowski2002Predictivecontrol}, such as road speed limits and minimum safe inter-vehicle distances.
The desired control actions from a constrained \ac{MPC} controller will not exceed a vehicle's performance limit or control vehicles into situations that could lead to an accident.
Additionally, so-called economic \ac{MPC} \cite{Amrit2011Economicoptimizationusing}, that assigns real values, such as fuel costs, to the weights in the cost function, has been utilized \cite{He2019Fuelefficiencyoriented} to link vehicle performance to an energy or financial metric.

Distributed \ac{MPC} algorithms in the leader-follower approach have been applied to platoons with poor communications \cite{Zhou2012VehiclePlatoonControl}, with extension to heterogeneous platoons \cite{Zheng2017DistributedModelPredictive}, and string stability was enforced using constraints \cite{Zhou2019Distributedmodelpredictive}.
A more complex approach was employed in \cite{Nunen2019StringStableModel} to include network information as a delay on the desired control action in the dynamics.

Most control designs for the platooning of vehicles consider a distributed approach with the use of a lead or ego vehicle that provides an input reference to the platoon.
This is a flexible approach as it allows for control designs to break apart and reform platoons \cite{Alam2015IEEEControlSystemsHeavyDutyVehicle}
However, distributed policies have been introduced that slow front vehicles and speed up later vehicles to form a platoon, while observing that this may be in conflict with the lead driver's goal of reaching their destination quickly \cite{Liang2016IEEETransactionsonIntelligentTransportationSystemsHeavyDutyVehicle}.
Additionally, \cite{Yanakiev1998IEEETransactionsonVehicularTechnologyNonlinearspacingpolicies} noted that the use of a variable headway and \ac{ACC} controller introduced a ``group conscience'' such that the leading vehicles were designed with reduced performance to take into account later vehicles in the platoon.
A centralized control design would utilize all platoon information and a platoon reference to design the control actions for the vehicles as a collection.

The original work on the control of vehicular platoons is \cite{Levine1966IEEETransactionsonAutomaticControloptimalerrorregulation}.
The authors designed a centralized \ac{LQR} controller that took a target reference velocity for the platoon and desired inter-vehicle spacing, to generate the control action for all vehicles in the platoon, which was furthered in \cite{Melzer1971AutomaticaOptimalregulationsystems}.
However, in \cite{Jovanovic2005IEEETransactionsonAutomaticControlillposednesscertain} it was shown that the original cost function in \cite{Levine1966IEEETransactionsonAutomaticControloptimalerrorregulation} is not string stable as the length of the platoon goes to infinity, such that as more vehicles are added the convergence time expands, and the initial control action increases.
The authors posed an alternative state representation and cost function that penalized both the absolute position error to the reference as well as the inter-vehicle distances to achieve finite convergence \cite{Jovanovic2005IEEETransactionsonAutomaticControlillposednesscertain}.

However, this approach is criticized in \cite{Curtain2010comparisonLQRcontrol} which shows that an infinite length platoon is not equivalent to a large but finite platoon.
In \cite{Jovanovic2004illposednesscertain} it is shown that the optimal control design fails for certain initial conditions with large control values resulting from the static gain computed from the \ac{LQR} such that desired control action could be larger than the maximum allowable control.
The poor performance of large platoons also occurs in decentralized designs where the state feedback control gain reduced for vehicles further away \cite{Lin2012OptimalControlVehicular}.
This reduction in state feedback gain was used to argue that for an $M$ length vehicle platoon, there should $M$ independent controllers with $M$ separately tuned gains \cite{Barooah2009MistuningBasedControl}.

%

\section{Platoon Architecture}
In this section, we state the single vehicle dynamics and develop the centralized platoon model of heterogeneous vehicles.
We then state our human driver model.

\subsection{Vehicle Dynamics}
We consider a commonly utilized linear dynamics for longitudinal motion of a vehicle-$i$.
Let us define our state variables as  $p^{(i)}$ [m] as a point at the front bumper, $v^{(i)}$ [m/s] the velocity, $a^{(i)}$ [m/s$^2$] acceleration, and $u^{(i)}$ [m/s$^2$] control input or desired acceleration.
Following \cite{rajamani2012vehicle}, the time derivative of each state is defined as  $\dot{p}^{(i)} = v^{(i)}$, $\dot{v}^{(i)} = a^{(i)}$, and
\begin{equation*}
    \dot{a}^{(i)} = -\frac{1}{\tau_i} a^{(i)} + \frac{1}{\tau_i} u^{(i)}
\end{equation*}
where  $\tau_i$ [s] is the mechanical actuation lag.

We write the state vector of a single vehicle-$i$ as $x^{(i)} = [p^{(i)}, v^{(i)}, a^{(i)}]^\mathsf{T}$,
which gives the standard state space form
\begin{equation}
    \dot{x}^{(i)} = A_c^{(i)} x^{(i)} + B_c^{(i)} u^{(i)},
    \label{eq:singlecontinuousdynamics}
\end{equation}
where $A_c^{(i)}$ and $B_c^{(i)}$ are the dynamics and control input matrices with the mechanical lag term for vehicle-$i$ and are given in Appendix~\ref{sec:app:singledynamics}.

Following \cite{AAstrom2011ComputerControlledSystems}, a continuous-time system \eqref{eq:singlecontinuousdynamics}, can be discretized in time with constant sampling interval (period) $\Delta_t$ [s] to
\begin{equation*}
 	x^{(i)}_{k + 1} = A^{(i)} x^{(i)}_k + B^{(i)} u^{(i)}_k + w^{(i)}_k
\end{equation*}
where the subscript $k$ is discrete-time, $w^{(i)}_k$ is i.i.d. process noise representing error in the discrete-time prediction model, modeled as zero mean normally distributed with covariance $\mathcal{W}^{(i)} > 0$, $w^{(i)}_k \sim \mathcal{N}(0,\mathcal{W}^{(i)})$ and the dynamics matrices are discretized using
\begin{equation*}
	A^{(i)} = \exp(A_c^{(i)} \Delta_t) ~ \textrm{and} ~ B^{(i)} = \int_0^{\Delta_t} \exp(A_c^{(i)} m) dm ~ B_c^{(i)}
\end{equation*}
and are given in Appendix~\ref{sec:app:singledynamics}.

Following \cite{Levine1966IEEETransactionsonAutomaticControloptimalerrorregulation,Melzer1971AutomaticaOptimalregulationsystems} and \cite{Jovanovic2005IEEETransactionsonAutomaticControlillposednesscertain}, we consider a combined model of the platoon.
For $M$ vehicles, we define the centralized multiple-output state and multiple-input control vectors as
\begin{align}
	X_k &= [p_k^{(1)}, \dots, p_k^{(M)}, v_k^{(1)}, \dots, v_k^{(M)}, a_k^{(1)}, \dots, a_k^{(M)}]^\mathsf{T}
	\label{eq:dynamics:platoonstate} \\
	U_k &= [u_k^{(1)}, \dots, u_k^{(M)}]^\mathsf{T}
	\label{eq:dynamics:platooncontrolvalue}
\end{align}
such that the platoon dynamics are
\begin{equation}
	X_{k+1} = A_M X_k + B_M U_k + W_k
	\label{eq:platoondynamics}
\end{equation}
where $A_M$ and $B_M$ are block diagonal matrices of the collection of single-vehicle dynamics and control input matrices, $W_k$ is a vector of the i.i.d. process noise acting on each vehicle which can be modeled as $W_k \sim \mathcal{N}(0,\mathcal{W})$.
The matrices $A_M$ and $B_M$ are given in Appendix~\ref{sec:app:platoondynamics}.

In \ac{MPC} design, instead of directly computing the platoon control action $U_k$, we optimize for the change in control actions $\Delta U_k$ from the previous control action such that the applied control \eqref{eq:dynamics:platooncontrolvalue} to the platoon dynamics \eqref{eq:platoondynamics} is
\begin{equation}
    U_k = U_{k-1} + \Delta U_k
    \label{eq:changeincontrol}
\end{equation}
where $\Delta U_k = [\Delta u_k^{(1)}, \dots, \Delta u_k^{(M)}]^\mathsf{T}$ and $\Delta u_k^{(i)}$ is the change in control action for vehicle-$i$.
This optimization is computed over a finite horizon of $N$ time steps into the future.
We use the platoon model \eqref{eq:platoondynamics} with \eqref{eq:changeincontrol} to predict the value of the state over the next $N$ time steps.
We introduce the predicted state value of the platoon at time $k+j$ for $j \in \{1, \dots, N\}$ from the measured state value at time $k$ using the model denoted as $\hat{X}_{k+j|k}$, with the prediction window defined as
\begin{equation*}
	\mathcal{X}_k = [\hat{X}_{k+1|k}^\mathsf{T}, \dots, \hat{X}_{k+N|k}^\mathsf{T}]^\mathsf{T}
\end{equation*}
for the predicted value of the change in control from the platoon controller as
\begin{equation*}
	\Delta \hat{U}_k = [\Delta \hat{U}_{k|k}^\mathsf{T}, \dots, \Delta \hat{U}_{k+N-1|k}^\mathsf{T}]^\mathsf{T}
\end{equation*}
where from the measurement at time $k$ the predicted applied control at time $k$ is $\hat{U}_{k|k} = U_{k-1} + \Delta \hat{U}_{k|k}$ and the predicted control at time $k+j$ is $\hat{U}_{k+j|k} = \hat{U}_{k+j-1|k} + \Delta \hat{U}_{k+j|k}$ for $j = \{1,\dots,N-1\}$.

We note that $\Delta U_{k}$ is the change in control applied at time $k$, while $\Delta \hat{U}_k$ is the predicted change in control over the finite horizon of length $N$.
The actual applied control action is not necessarily equal to the prediction.

Using algebraic manipulation as illustrated in \ac{MPC} texts (e.g. \cite{Maciejowski2002Predictivecontrol}) the state prediction of the platoon $\mathcal{X}_k$ can be written as a linear combination of the current state $X_k$, the previous applied control $U_{k-1}$ and the predicted change in control $\Delta \hat{U}_k$
\begin{equation}
	\mathcal{X}_k = \Phi X_k + \lambda U_{k-1} + \Gamma \Delta \hat{U}_k
	\label{eq:mpc:predictionmodel}
\end{equation}
where $\Phi$ is the propagation of the state through the dynamics matrix $A_M$, $\lambda$ and $\Gamma$ are the propagation of the control inputs through the dynamics and control matrices, and are given in Appendix~\ref{sec:app:platoondynamics}.

We utilize this state prediction model to design a centralized \ac{MPC} for the coordinated control of a platoon of vehicles.

\subsection{Human Driver Model}
We consider scenarios where during operation of the platoon a human driver temporarily takes control of their vehicle.
This raises the critical challenge that our platoon controller must reconfigure to this human driver to ensure safety of the platoon.
A human driver may choose to modify their interaction with the \ac{CACC} automation due to traffic conditions or for safety reasons \cite{Dey2016IEEETransactionsonIntelligentTransportationSystemsReviewCommunicationDriver}.
After switching to manual operation, we make no assumption about the human driver's decision to maintain a new velocity, or inter-vehicle distance.
While algorithmic methods could be utilized to produce a prediction of the human driver actions, such as economic \ac{MPC} \cite{Fleming2021IncorporatingDriverPreferences} or human behavior \cite{Dey2016IEEETransactionsonIntelligentTransportationSystemsReviewCommunicationDriver}, these methods would require an assumption or model of the intended human driver action.
For our \ac{CACC} design, we make limited assumptions.
We assume that the human driver is solely focused on the state of their own vehicle, does not interact with any other vehicles in the platoon, and issues control actions that are consistent with physical (engine limit) and legal (road speed limit) constraints.

Consider a vehicle-$\ell$ has temporarily left the platoon and has the change in control action $\Delta \mathfrak{u}_k^{(\ell)}$ from the human driver replacing the platoon control $\Delta u_k^{(\ell)}$ such that the control action is $u_k^{(\ell)} = u_{k-1}^{(\ell)} + \Delta \mathfrak{u}_k^{(\ell)}$.
For ease of notation we modify the platoon change in control action \eqref{eq:changeincontrol} with a switch
\begin{equation}
    U_k = U_{k-1} + \alpha_k \Delta U_k + \bar{\alpha}_k \Delta \mathcal{U}_k
    \label{eq:platoonchangeincontrolwithhuman}
\end{equation}
where $\Delta \mathcal{U}_k$ is the control action applied from a human driver, a binary switch $\alpha_k$ as a diagonal square matrix of size $M$ that takes ones on the diagonal for the vehicles controlled by the platoon and zero in the $i,i$th element when vehicle-$i$ is not controlled by the platoon controller, and $\bar{\alpha}_k = I_M - \alpha_k$.
When the platoon is fully controlled by the centralized platoon controller $\alpha_k \equiv I_M$ and $\bar{\alpha}_k \equiv 0_M$, and \eqref{eq:platoonchangeincontrolwithhuman} reduces to \eqref{eq:changeincontrol}.
The dynamics of the platoon \eqref{eq:platoondynamics} are now
\begin{equation}
    X_{k+1} = A_M X_k + B_M U_{k-1} + B_M \alpha_k \Delta U_k + B_M \bar{\alpha}_k \Delta \mathcal{U}_k .
    \label{eq:platoonmodel}
\end{equation}

Based on the applied control at time $k-1$, the platoon controller is aware if every vehicle has utilized the centralized platoon controller or an alternative control value.
As such, $\alpha_k$ is known to the controller at time $k$.
If a vehicle has temporarily left the platoon, we assume that the vehicle will continue to be human controlled until informed otherwise, and $\alpha_k$ is constant for the finite prediction horizon.
The finite horizon prediction for the state of the platoon is expanded from  \eqref{eq:mpc:predictionmodel} to
\begin{equation}
	\mathcal{X}_k = \Phi X_k + \lambda U_{k-1} + \Gamma (I_N \otimes \alpha_k) \Delta \hat{U}_k + \Gamma (I_N \otimes \bar{\alpha}_k) \Delta \hat{\mathcal{U}}_k
	\label{eq:mpc:predictionmodelfull}
\end{equation}
where $ \Delta \hat{\mathcal{U}}_k = [\Delta \hat{\mathcal{U}}_{k|k}^\mathsf{T}, \dots, \Delta \hat{\mathcal{U}}_{k+N-1|k}^\mathsf{T}]^\mathsf{T}$ are future change in controls from the human driver,
$\otimes$ is the Kronecker product and $I_N$ is the identity matrix of size $N$.

In the following, we design a finite horizon cost function to find the optimal change in control $\Delta U_k$ for the platoon, which requires knowledge of any human driver control action $\Delta \mathcal{U}_k$.
Ideally for the platoon controller, the future human driver change in control actions $\Delta \mathcal{U}_k$ are known exactly, however, this is unlikely to be the case.
This motivates the use of a predicted control action for the human driver control values.
For a finite prediction horizon of length $N$, the human-driver model can be written as
\begin{equation}
	\mathcal{X}_k^{(\ell)} = \bar{\Phi} x_k^{(\ell)} + \bar{\lambda} u_{k-1}^{(\ell)} + \bar{\Gamma} \Delta \hat{\mathcal{U}}_k^{(\ell)} ,
	\label{eq:humandriverdynamics}
\end{equation}
where $\bar{\Phi}$, $\bar{\lambda}$ and $\bar{\Gamma}$ are given in Appendix~\ref{sec:app:platoondynamics} and
$\Delta \hat{\mathcal{U}}_k^{(\ell)} = [\Delta \hat{\mathfrak{u}}_{k+1|k}^{(\ell)}, \dots, \Delta \hat{\mathfrak{u}}_{k+N-1|k}^{(\ell)}]^\mathsf{T}$ is the prediction of the human driver change in control of vehicle-$\ell$.

We utilize \eqref{eq:humandriverdynamics} to compute a basic prediction of the human driver's control action, which we can utilize in the platoon prediction model \eqref{eq:mpc:predictionmodelfull} to design a centralized \ac{MPC} for the coordinated control of a platoon of vehicles.

%

\section{Platooning Problem}%
\label{sec:controlprob}
We desire to control the entire platoon to reach a target velocity of $v_{d}$ [m/s] with the desired distance between vehicle-$i$ and its immediate predecessor vehicle-$(i-1)$ as
\begin{equation}
\bar{d}^{(i)}_k \triangleq d_i + h_k^{(i)} v_k^{(i)} = l_{i-1} + r_{i} + h_k^{(i)} v_k^{(i)}
	\label{eq:singlevehicledesireddistance}
\end{equation}
where $d_i = l_{i-1} + r_i$ is the constant inter-vehicle distance, $l_{i-1}$ [m] is the length of vehicle-$(i-1)$, $r_{i}$ [m] the desired standstill distance in front of vehicle-$i$, and $h_k^{(i)}$ [s] the desired headway time.
The headway time quantifies the distance to the preceding vehicle at the current velocity.
The desired standstill distance $r_i$ and headway time $h_k^{(i)}$ are vehicle specific and can be chosen by the respective driver, whereas the desired velocity is platoon specified.

We consider a unique headway time for each vehicle, which can be modified by the occupants of the vehicle.
Commercially available \ac{ACC} systems allow for user selection of headway time \cite{Wu2020IEEEAccessSpacingPoliciesAdaptive}, with increments at $1$, $1.5$ and $2$ seconds \cite{Xiao2011IEEETransactionsonIntelligentTransportationSystemsPracticalStringStability}.
We write the individual headways, $h_k^{(i)}$, as a function of time $k$, to indicate that these can be modified but consider that a reasonable driver would not be constantly changing their headway.

Additionally, we desire to ensure the following constraints for all vehicles $i \in \{1,M\}$
\begin{itemize}
	\item[C1:] $p^{(i-1)} - p^{(i)} \geq d_{\textrm{min}}$, minimum safe distance between vehicles to ensure that no vehicle impacts its predecessor,
	\item[C2:] $p^{(i-1)} - p^{(i)} \leq d_{\textrm{max}}$ maximum distance between vehicles to ensure (random) communications are maintained,
	\item[C3:] $v_{\textrm{min}} \leq v^{(i)}$, minimum velocity set to zero on the assumption that no vehicle in the platoon reverses on the road,
	\item[C4:] $v^{(i)} \leq v_{\textrm{max}}$, maximum velocity chosen based on the road speed limit, or the performance limitation of a vehicle,
	\item[C5:] $a_{\textrm{min}} \leq a^{(i)}$, minimum acceleration bounded based on the performance of the braking systems, and
	\item[C6:] $a^{(i)} \leq a_{\textrm{max}}$, maximum acceleration chosen based on the engine performance of the vehicles.
\end{itemize}
The acceleration bounds could be further limited for the comfort of the vehicle occupants.

Finally, we also consider that our proposed controller can accommodate a human driver taking temporary control of their vehicle within the platoon.
This could include a driver initiating an emergency brake, reducing speed, or temporarily maintaining a larger distance from the previous vehicle than specified.
This accommodation allows for a human driver to drive within the bounds of the platoon to their own comfort.
Additionally, it may allow for the inclusion of legacy vehicles.
We make the minimum assumption that the vehicle and driver will obey performance limits of the vehicle: the minimum and maximum accelerations, and legal limits on velocity: non-negative and not exceeding the road speed limit.

We consider the situation where it is more important for the platoon to stay together, and an emergency brake for one vehicle should be obeyed by the platoon.
This is in contrast to control policies in \cite{Alam2015Experimentalevaluationdecentralized,Alam2015IEEEControlSystemsHeavyDutyVehicle} where each vehicle has individual goals and platoons are allowed to split and reform.
We desire to ensure that our proposed controller yields a stable closed-loop to temporary inputs from a human driver to their individual vehicle within the minimum constraints.

We finish our platooning problem by stating our problem in the form of an optimization problem:
\begin{equation*}
    \min \sum_{k=0}^\infty \sum_{i=1}^M (v_k^{(i)} - v_d)^2 + (p_k^{(i-1)} - p_k^{(i)} - \bar{d}_k^{(i)})^2 + (\Delta \hat{U}^{(i)})^2
\end{equation*}
subject to: \eqref{eq:platoonmodel}, \eqref{eq:humandriverdynamics}, and constraints C1--6.

First, the inclusion of the system and safety constraints necessitates a careful reference design.
As demonstrated in \cite{Quevedo2012IEEETransactionsonPowerElectronicsModelPredictiveControl}, careful design of reference signals may be required to ensure that computed control actions smoothly converge the states to the desired values while not violating system dynamics and constraints.
We present a suitable time-varying reference in Section~\ref{subsec:referencedesign}.

Second, it is not possible to solve the infinite horizon cost function due to the required computations to ensure constraints and potential interruption from a human driver.
However, through careful design of the terminal cost in a finite horizon cost function, the infinite horizon design can be recovered \cite{Scokaert1998Constrainedlinearquadratic}.
We present our platoon cost function and optimization in Section~\ref{subsec:costfuncdesign}.

Third, the future states of the human driver vehicle are required for the prediction of the platoon state, see \eqref{eq:mpc:predictionmodelfull}. 
To predict the state of the human driver vehicle, we predict the potential control actions of the human driver by posing a limited assumption human driver cost function.
We present the human driver control action prediction in Section~\ref{subsec:humandrivermodel}.

Finally, we summarize our control design with an algorithmic implementation in Section~\ref{subsec:controllerimplementation}.

%

\section{Controller Design}

We now design our controller using the models given above to achieve the desired platoon velocity and inter-vehicle distances while guaranteeing the constraints and able to reconfigure to a temporary human driver.
First, we design a time-varying reference for the platoon.
Second, we design our finite horizon cost function for the platoon inspired by the infinite horizon cost function of \cite{Jovanovic2005IEEETransactionsonAutomaticControlillposednesscertain}.
We also apply the desired constraints on the cost function to propose our constrained \ac{MPC} controller to centrally control the platoon to the desired platoon velocity and inter-vehicle distances.
Third, we propose a simple finite horizon cost function to predict the human driver control actions for use in the platoon controller.
Finally, we summarize the implementation of our platoon control design with human driver interaction in an algorithm.

\subsection{Reference Design}%
\label{subsec:referencedesign}

Proportional state feedback controllers have used in several platooning works \cite{rajamani2012vehicle}, including \ac{LQR} \cite{Levine1966IEEETransactionsonAutomaticControloptimalerrorregulation,Melzer1971AutomaticaOptimalregulationsystems,Jovanovic2005IEEETransactionsonAutomaticControlillposednesscertain,Lin2012OptimalControlVehicular,Alam2015Experimentalevaluationdecentralized}.
However, for constant gain feedback regulators, the control value increases the further the states are from the desired reference \cite{Anderson1990OptimalControlLinear}.
In reasonable platooning scenarios \cite{Jovanovic2004illposednesscertain}, such as zero initial velocity, the desired initial control actions could exceed maximum allowable control action \cite{Jovanovic2004avoidingsaturationcontrol}.

To avoid this issue with constant static references used in regulators, we propose a time-varying reference for the desired platoon states.
Using a slowly increasing reference, all vehicles in the platoon are able to converge to the desired reference before the position and velocity references reach the desired steady-state.
This allows convergence to the reference from any initial condition.

We consider a slowly increasing ramp for the velocity reference with constant acceleration from initial time $k_0$ as
\begin{equation*}
	v_k^\star =
	\begin{cases}
		a_k^\star \Delta_t k + \bar{v}, \quad k_0 \leq k < k_0 + k_m \\
		v_d, \quad k \geq k_0 + k_m
	\end{cases}
\end{equation*}
where $\bar{v} = \min v_{k_0}^{(i)}$ is initialized to the minimum velocity of the platoon, the acceleration reference is
\begin{equation*}
	a_k^\star =
	\begin{cases}
		\frac{v_d - \bar{v}}{\Delta_t k_m}, \quad k_0 \leq k < k_0 + k_m \\
		0, \quad k \geq k_0 + k_m
	\end{cases}
\end{equation*}
and $k_m$ is the sampling periods to reach the desired velocity.
The time constant $k_m$ is a tuning parameter of the controller.
We include acceleration in our reference design as it is necessary to provide a reference for all states.
It is demonstrated in \cite{Quevedo2012IEEETransactionsonPowerElectronicsModelPredictiveControl} that it is important to design a reference for all states, such that the reference satisfies both desired constraints and system dynamics, while providing a smooth path to the final desired values.

For the position reference, we take inspiration from \cite{Melzer1971AutomaticaOptimalregulationsystems} and \cite{Jovanovic2005IEEETransactionsonAutomaticControlillposednesscertain} to establish the position reference of all the vehicles as the cumulative sum of the desired distances from a virtual lead vehicle-$0$.
The lead vehicle position reference is
\begin{equation*}
	p_k^\star = \begin{cases}
		\frac{1}{2} a_k^\star (\Delta_t k)^2 + \bar{v} \Delta_t k + \bar{p}, \quad k_0 \leq k < k_0 + k_m \\
		v_d \Delta_t k + \bar{p}, \quad k \geq k_0 + k_m
	\end{cases}
\end{equation*}
where $\bar{p} = p^{(1)}_{k_0} + \bar{d}_{k_0}^{(1)}$ is initialized from the position of vehicle-$1$.
The position reference for each vehicle-$i$ is
\begin{equation*}
	p_k^{(i)\star} = p_k^\star - \sum_{j=1}^i \bar{d}_k^{(j)} = p_k^\star - \left(\sum_{j=1}^i d_j + h_k^{(j)} v_k^\star \right) ,
\end{equation*}
where $\bar{d}_k^{(i)}$ is defined in \eqref{eq:singlevehicledesireddistance}.
By using the desired inter-vehicle distanced to form the position referenced for each individual vehicle, the headway times are included as part of the state reference.

When a vehicle leaves the platoon under human driver control, we desire to drive the platoon forward at the desired velocity but within the platoon constraints. 
We reset the platoon reference based on the human controlled vehicle state.
The initial time is set as $k_0 = k$, and velocity reference is set to the velocity of  vehicle-$\ell$: $\bar{v} = v_{k_0}^{(\ell)}$, and the virtual lead vehicle position as the desired distance from vehicle-$\ell$: $\bar{p} = p_k^{(\ell)} + \sum_{j=1}^\ell \bar{d}_k^{(j)}$.

For convenience we define our desired reference for the platoon at time $k$ as the vector $X_k^\star = [p_k^{(1)\star}, \dots, p_k^{(M)\star}, v_k^{\star}, \dots, v_k^{\star}, a_k^{\star}, \dots, a_k^{\star}]^\mathsf{T}$
and over the finite prediction horizon as $\mathcal{X}_k^\star = [(X_{k+1}^\star)^\mathsf{T}, \dots, (X_{k+N}^\star)^\mathsf{T}]^\mathsf{T}$.

\subsection{Cost Function Design}%
\label{subsec:costfuncdesign}

To design our \ac{MPC} platoon controller we establish position, velocity and acceleration error states using our time-varying references.
We propose a finite horizon cost function of these errors and discuss how our cost function can be rearranged to be in a quadratic function of the vehicle states.
Finally, we apply the desired inter-vehicle distance, velocity, and acceleration limits as state constraints on the cost function.
Our final constrained cost function is in the form of a quadratic program, which can then be solved using standard convex optimization techniques.
The constraints on the states and control are a boundary in the cost function solution space, such that the predicted optimal control action is guaranteed to not exceed the desired constraints.
There exist several quadratic programming solvers to establish the optimal control action within constraints \cite{Maciejowski2002Predictivecontrol} which reduces to optimization of a convex function \cite{Boyd2004ConvexOptimization}.

Consider for each vehicle-$i$ for $i \in \{1,M\}$, the absolute position, velocity, and acceleration errors as the difference between the current state and desired reference
	$\xi^{(i)}_k = p^{(i)}_k - p_k^{(i) \star}$,
	$\zeta^{(i)}_k = v^{(i)}_k - v_k^{\star}$, and
	$\psi_k^{(i)} = a_k^{(i)} - a_k^{\star}$.
For the entire platoon, these errors can be written as $X_k - X_k^\star = [\xi^{(1)}_k, \dots, \xi^{(M)}_k, \zeta^{(1)}_k, \dots, \zeta^{(M)}_k, \psi_k^{(1)}, \dots, \psi_k^{(M)}]$.
For convenience below, we define $\hat{\eta}_{k+j|k}^{(i)}$, $\hat{\xi}_{k+j|k}^{(i)}$,  $\hat{\zeta}_{k+j|k}^{(i)}$, and $\hat{\psi}_{k+j|k}^{(i)}$ as the predicted errors where the subscript indicates the state prediction at time $k+j$ given the state at time $k$.

Following \cite{Melzer1971AutomaticaOptimalregulationsystems} and \cite{Jovanovic2005IEEETransactionsonAutomaticControlillposednesscertain}, we introduce virtual reference vehicles on the platoon boundary that perfectly follow the reference $p_k^{(0)} = p_k^\star$, $p_k^{(M+1)} = p_k^{(M+1)\star}$, $v_k^{(0)} = v_k^{(M+1)} = v_k^\star$, and $a_k^{(0)} = a_k^{(M+1)} = a_k^\star$.
and introduce the relative position error between vehicle-$i$ and vehicle-$(i-1)$ for $i \in \{1,M+1\}$ as
\begin{equation}
	\eta^{(i)}_k = p^{(i)}_k - p_k^{(i-1)} + \bar{d}_k^{(i)} .
	\label{eq:relativeposerror}
\end{equation}

Inspired by the infinite horizon cost function of \cite{Jovanovic2005IEEETransactionsonAutomaticControlillposednesscertain} we propose a finite horizon cost function over a prediction horizon of $N$ steps with our time-varying references
\begin{align}
	J &= \sum_{j=0}^{N-1} \left[ \sum_{i=1}^{M+1} q_1 \left(\hat{\eta}_{k+j|k}^{(i)}\right)^2 + \sum_{i=1}^M \left(q_2 \left(\hat{\xi}_{k+j|k}^{(i)}\right)^2 \right.\right.\nonumber\\&\quad\left.\left.+ q_3 \left(\hat{\zeta}_{k+j|k}^{(i)}\right)^2 + q_4 \left(\hat{\psi}_{k+j|k}^{(i)}\right)^2 + r \left(\Delta u_{k+j|k}^{(i)}\right)^2 \right) \right] \nonumber
	\\&\quad+ (\hat{X}_{k+N|k} - X_{k+N}^\star)^\mathsf{T} P_{k+N} (\hat{X}_{k+N|k} - X_{k+N}^\star)
	\label{eq:costfuncwithheadway}
\end{align}
where
$q_1$ is the penalty on relative position error, $q_2$ the penalty on absolute position error, $q_3$ the penalty on velocity error, $q_4$ as the penalty on the acceleration, $r$ the penalty on the control inputs, and $P_{k+N}$ is the terminal state cost.
To achieve convergence independent of platoon length, it is necessary to penalize both the relative and the absolute position errors \cite{Jovanovic2005IEEETransactionsonAutomaticControlillposednesscertain}.

Using algebraic manipulation and \eqref{eq:singlevehicledesireddistance}, the relative position error \eqref{eq:relativeposerror}, can be written as a function of the errors
\begin{align*}
	\eta^{(i)}_k
	= &\xi_k^{(i)} - \xi_k^{(i-1)} + h_k^{(i)} \zeta_k^{(i)}
\end{align*}
such that the relative position errors can be incorporated as cross-terms of the absolute position errors and velocity errors, with the headway times as a weight on the velocity errors.
While one could think of the headway times as a reference to the problem as introduced in the desired inter-vehicle distance \eqref{eq:singlevehicledesireddistance}, it is more convenient as a weight on the state deviation.
By forcing the headway time to be a state reference, it may lead to a nonlinear control problem.

Our cost function can now be efficiently written as a quadratic function
\begin{align}
	 &J = (\hat{X}_{k+N|k} - X_{k+N}^\star)^\mathsf{T} P_{k+N} (\hat{X}_{k+N|k} - X_{k+N}^\star) \label{eq:condensedcostfunc} \\
	 &+ \sum_{j=0}^{N-1} \left( (\tilde{X}_{k+j|k})^\mathsf{T} Q_{k+j} \tilde{X}_{k+j|k} + \Delta \hat{U}_{k+j|k}^\mathsf{T} R_\Delta \Delta \hat{U}_{k+j|k} \right) \nonumber
\end{align}
where $\tilde{X}_{k+j|k} = \hat{X}_{k+j|k} - X_{k+j}^\star$, $R_\Delta = r I_M$, and
\begin{equation}
	Q_\kappa = \begin{bmatrix} q_1 T_M + q_2 I_M & q_1 T_\kappa & 0 \\ q_1 T_\kappa^\mathsf{T} & q_1 H_\kappa + q_3 I_M & 0 \\ 0 & 0 & q_4 I_M \end{bmatrix}
	\label{eq:beststatepenalty}
\end{equation}
where $0$ is a square matrix of zeros of size $M \times M$, $T_M$ is a symmetric Toeplitz matrix of size $M \times M$ with the first row of the form $[2, -1, 0, \dots, 0]$,  $T_\kappa$ is an $M \times M$ matrix with the headway times of all vehicles $[h_\kappa^{(1)}, \dots, h_\kappa^{(M)}]$ on the diagonal and negative headway times of vehicles-$2$ to-$M$ on the first upper diagonal $[-h_\kappa^{(2)}, \dots, -h_\kappa^{(M)}]$,
and $H_\kappa$ is a diagonal $M \times M$ matrix where
$H_\kappa = \textrm{diag} [(h_\kappa^{(1)})^2, \dots, (h_\kappa^{(M)})^2 ]$.
In the case of a common constant time headway across the platoon $h_k^{(i)} = h$, then $T_\kappa$ reduces to a Toeplitz matrix with $h$ on the diagonal and $-h$ on the first upper diagonal, and $H_\kappa$ reduces to the identity $I_M$ multiplied by $h^2$.
The reduction of \eqref{eq:costfuncwithheadway} to \eqref{eq:condensedcostfunc} is given in Appendix~\ref{sec:app:costfunc}.

The terminal penalty $P$ is the penalty on the final state in the prediction horizon.
Choosing $P$ as the solution of the algebraic Ricatti equation implements the infinite horizon cost on the final state such that the final control action is the infinite horizon optimal control action \cite{Scokaert1998Constrainedlinearquadratic}.

Using algebraic manipulation and \eqref{eq:mpc:predictionmodelfull} the cost function can be written in the form of a quadratic program
\begin{align}
    &J(X_k, \Delta \hat{U}_k) = f(X_k, U_{k-1}) \label{eq:quadraticprogramform} \\ &\quad+ \Delta U_k^\mathsf{T} (I_N \otimes \alpha_k)^\mathsf{T} (\Psi + \Gamma^\mathsf{T} \Omega \Gamma) (I_N \otimes \alpha_k) \Delta U_k \nonumber \\ &\quad+ 2(\Phi X_k + \lambda U_{k-1} + \Gamma (I_N \otimes \bar{\alpha}_k) \Delta \hat{\mathcal{U}}_k - \mathcal{X}_k^\star)^\mathsf{T} \nonumber \\ &\quad\times \Omega \Gamma (I_N \otimes \alpha_k) \Delta \hat{U}_k \nonumber
\end{align}
where $f(X_k, U_{k-1})$ is a constant term and
	$\Omega = \textrm{diag}\{Q_k,\dots,Q_{k+N-1},P_{k+N}\},$
	$\Psi = \textrm{diag}\{R_\Delta, \dots, R_\Delta\}$
are block diagonal matrices.
Thus the optimization problem can be solved using a standard quadratic program solver.

Consider the desired constraints on the vehicles' velocities and accelerations, and the inter-vehicle distances outlined in Section~\ref{sec:controlprob} of inter-vehicle distance $d_{\textrm{min}} \leq p_k^{(i-1)} - p_k^{(i)} \leq d_{\textrm{max}}$, velocity $v_{\textrm{min}} \leq v_k^{(i)} \leq v_{\textrm{max}}$, and acceleration $a_{\textrm{min}} \leq a_k^{(i)} \leq a_{\textrm{max}}$
which can be written as a matrix inequality of the platoon state
\begin{equation*}
	\begin{bmatrix} \check{G} & g \end{bmatrix} \begin{bmatrix} \hat{X}_{k+j|k} \\ 1 \end{bmatrix} \leq 0
\end{equation*}
where
\begin{equation*}
	\begin{bmatrix} \check{G} & g \end{bmatrix} =
	\begin{bmatrix}
		\mathfrak{T}_M & 0 & 0 & 1_{M-1} d_{\textrm{min}} \\
		-\mathfrak{T}_M & 0 & 0 & -1_{M-1} d_{\textrm{max}} \\
		0 & -I_M & 0 & 1_{M} v_{\textrm{min}} \\
		0 & I_M & 0 & -1_{M} v_{\textrm{max}} \\
		0 & 0 & -I_M & 1_{M} a_{\textrm{min}} \\
		0 & 0 & I_M & -1_{M} a_{\textrm{max}}
	\end{bmatrix}
\end{equation*}
where $\mathfrak{T}_M $ is a size $(M-1) \times M$ Toeplitz matrix with $-1$ on the diagonal and $1$ on the first upper diagonal, and $1_{M-1}$ and $1_{M}$ are column vectors of ones of size $(M-1)$ and $M$, respectively, for a total of $6M-2$ constraints for each step of the prediction horizon.

These constraints can be extended over the finite prediction horizon
\begin{equation*}
    \begin{bmatrix} \bar{G} & \bar{g} \end{bmatrix} \begin{bmatrix} \mathcal{X}_k \\ 1 \end{bmatrix} \leq 0
\end{equation*}
where $\bar{G} = \textrm{diag} [\check{G}, \dots, \check{G}]$ and $\bar{g}^\mathsf{T} = [g^\mathsf{T}, \dots, g^\mathsf{T}]$.
Using the prediction model \eqref{eq:mpc:predictionmodelfull}, the constraints on the states can be written in terms of $\Delta \hat{U}_k$
\begin{align}
    \bar{G} &\Gamma (I_N \otimes \alpha_k) \Delta \hat{U}_k \nonumber \\ &\leq - \bar{G} \left(\Phi X_k + \lambda U_{k-1} + \Gamma (I_N \otimes \bar{\alpha}_k) \Delta \hat{\mathcal{U}}_k\right) - \bar{g}
    \label{eq:qpconstraints}
\end{align}
such that the state constraints appear as a boundary on the cost function \eqref{eq:quadraticprogramform} \cite{Maciejowski2002Predictivecontrol}.

The left of \eqref{eq:qpconstraints} only changes in the event a vehicle leaves the platoon and $\alpha_k \neq \alpha_{k-1}$.
For a finite platoon of $M$ vehicles, it is possible to pre-compute all variations of $\bar{G} \Gamma (I_N \otimes \alpha_k)$.
However, the right of \eqref{eq:qpconstraints} is dependent on the current platoon state $X_k$ and the last control action $U_{k-1}$, and must be recomputed each step.

For a prediction horizon of length $N$, there are $N\times (6M-2)$ constraints across the $3M$ states at each time point, such that $\bar{G} $ is of size $N(6M-2) \times 3MN$.
Increasing the length of the prediction horizon, $N$, increases the number of control values required by the number of vehicles $M$ but the number of applied constraints by $6M-2$.
For a large platoon this can be a significant increase to the computational effort.

The optimal platoon control action is the change in control that minimizes the constrained finite horizon cost function
\begin{equation}
    \Delta \hat{U}_k^\star = \min_{\Delta \hat{U}_k} J(X_k, \Delta \hat{U}_k)
    \label{eq:platooncontrolminimisation}
\end{equation}
which we then implement the first element $\Delta \hat{U}_{k|k}^\star$ to the platoon, before solving again at the next time step.

\subsection{Incorporation of Hybridized Human Driver Model}%
\label{subsec:humandrivermodel}

We now predict the change in control action from a human driver for use in our platoon controller using a second \ac{MPC} algorithm.
We assume that the human driver will behave reasonably by rarely changing their control action, and will at minimum obey performance (acceleration) constraints of the vehicle, maintain non-negative velocity and obey the road speed limit.
The same acceleration and velocity assumptions are applied in our centralized platoon controller.
Future change in control actions from the human driver that violate these constraints could cause our platoon controller to be infeasible.

Using our assumption that the human driver will rarely change their control action, we consider that the driver's control actions will be constant over the finite horizon prediction time of the platoon controller.
We propose the following quadratic finite horizon cost function of the human driver
\begin{align*}
	&\bar{J}(x_k^{(\ell)}, \Delta \hat{\mathcal{U}}_k^{(\ell)}) = (\hat{x}^{(\ell)}_{k+N|k})^\mathsf{T} \bar{P} (\hat{x}^{(\ell)}_{k+N|k}) \nonumber \\ &+ \sum_{j=0}^{N-1} \left( (\hat{x}^{(\ell)}_{k+j|k})^\mathsf{T} \bar{Q} (\hat{x}^{(\ell)}_{k+j|k}) + (\Delta \hat{\mathfrak{u}}^{(\ell)}_{k+j|k})^\mathsf{T} r_\Delta (\Delta \hat{\mathfrak{u}}^{(\ell)}_{k+j|k}) \right)
\end{align*}
As we make no assumption on desired state of the human driver, we choose no penalty on the state such that $\bar{Q} = \bar{P} = 0$, and the penalty on control action as the same in the platoon model, where $r_\Delta$ is the $\ell$th diagonal element of $R_\Delta$.

This is the most uninformative cost function possible as it assumes the driver will make no changes to their current control value.
Our cost function simplifies to
\begin{equation}
	\bar{J}(x_k^{(\ell)}, \Delta \hat{\mathcal{U}}_k^{(\ell)}) = (\Delta \hat{\mathcal{U}}_k^{(\ell)})^\mathsf{T} \bar{\Psi} \Delta \hat{\mathcal{U}}_k^{(\ell)} \label{eq:humanmpc:costfunc}
\end{equation}
where $\bar{\Psi} = \textrm{diag}\{r_\Delta, \dots, r_\Delta \}$.
Clearly, this cost function is minimized when $\Delta \hat{\mathcal{U}}_k^{(\ell)} = 0$.

Now we consider the minimum constraints that we assume the human driver obeys of velocity $v_{\textrm{min}} \leq v_k^{(\ell)} \leq v_{\textrm{max}}$, and acceleration $a_{\textrm{min}} \leq a_k^{(\ell)} \leq a_{\textrm{max}}$ which can be written as the matrix inequality on the vehicle state at time $k+j$ as
\begin{equation*}
	\begin{bmatrix} \check{G} & g \end{bmatrix} \begin{bmatrix} \hat{x}_{k+j|k}^{(\ell)} \\ 1 \end{bmatrix} \leq 0
\end{equation*}
where
\begin{equation*}
	\begin{bmatrix} \check{G} & g \end{bmatrix} = \begin{bmatrix}
	0 & -1 & 0 & v_{\textrm{min}} \\
	0 & 1 & 0 & -v_{\textrm{max}} \\
	0 & 0 & -1 & a_{\textrm{min}} \\
	0 & 0 & 1 & -a_{\textrm{max}}
	\end{bmatrix} .
\end{equation*}
These constraints can be extended over the finite prediction horizon
\begin{equation*}
    \begin{bmatrix}
    \bar{G} & \bar{g}
    \end{bmatrix} \begin{bmatrix}
    \mathcal{X}_k^{(\ell)} \\ 1
    \end{bmatrix} \leq 0
\end{equation*}
where $\bar{G} = \textrm{diag} [\check{G}, \dots, \check{G}]$ and $\bar{g}^\mathsf{T} = [g^\mathsf{T}, \dots, g^\mathsf{T}]$.
Using algebraic manipulation with the vehicle dynamics \eqref{eq:humandriverdynamics}, the state constraints can be written as a function of the change in control $\Delta \hat{\mathcal{U}}_k^{(\ell)}$
\begin{equation}
    \bar{G} \bar{\Gamma} \Delta \hat{\mathcal{U}}_k^{(\ell)} \leq -\bar{G} \left(\bar{\Phi} x_k^{(\ell)} + \bar{\lambda} u_{k-1}^{(\ell)} \right) - \bar{g} .
    \label{eq:humanmpc:constraints}
\end{equation}

The quadratic cost function \eqref{eq:humanmpc:costfunc} with the linear matrix constraints \eqref{eq:humanmpc:constraints} can be minimized using standard quadratic programming solvers to find the minimum control action that meets the constraints.
We would only expect to predict a change in control action when one of the constraints will be violated in the finite horizon.
We take the prediction of the constrained but minimally penalized control for the human controlled vehicle-$\ell$, $\Delta \hat{\mathcal{U}}_k^{(\ell)}$, and include this in the computation for the centralized platoon control action.

\subsection{Controller Implementation}%
\label{subsec:controllerimplementation}

We now summarize our proposed control design with human driver interaction.
In Algorithm~\ref{algo:platoonimplementation} we outline the implementation of our centralized platoon \ac{MPC} with the human driver prediction.
The inputs to the controller are the number of vehicles, dynamics (mechanical actuation lags), desired headway times, inter-vehicle distances, platoon velocity, constraints, reference sampling periods for acceleration, prediction window length, and cost function weights.
In normal operation, the state reference is computed, the platoon change in control is optimized, the latest control action is computed, and used to update the dynamics.
At the start of each time-step, a check is performed to identify if a vehicle has left the platoon.
If a vehicle has left the platoon, then the reference parameters are updated based on the current state of the human driver controlled vehicle.
This is used to produce a prediction of the human drivers action.

\begin{algorithm}
 \caption{Centralized platoon control with human-driver interaction}
 \label{algo:platoonimplementation}
 \begin{algorithmic}[1]
 \renewcommand{\algorithmicrequire}{\textbf{Input:}}
 \renewcommand{\algorithmicensure}{\textbf{Output:}}
 \REQUIRE
 Number of vehicles $M$, \\
 Mechanical actuation lags $\tau_i$, \\
 Headway times $h_k^{(i)}$ and distances $r_i$, \\
 Desired velocity $v_d$, \\
 Constraints $\{d_\textrm{min}, d_\textrm{max}, v_{\textrm{min}}, v_{\textrm{max}}, a_{\textrm{min}}, a_{\textrm{max}}\}$, \\
 Reference sampling periods $k_m$, \\
 Prediction window $N$, and \\
 Cost function weights $\{q_1, q_2, q_3, q_4, r\}$.
  \FOR {time steps $k = 0, 1, \dots$}
  \IF {a vehicle has left the platoon}
    \STATE Identify human driver controlled vehicle-$\ell$
    \STATE Re-set state reference based on vehicle-$\ell$ position and velocity (Section~\ref{subsec:referencedesign})
    \STATE Predict change in human driver action $\Delta \hat{\mathcal{U}}_k^{(\ell)}$ by minimizing \eqref{eq:humanmpc:costfunc} subject to \eqref{eq:humanmpc:constraints}
  \ENDIF
  \STATE Compute state reference $\chi_k^\star$ (Section~\ref{subsec:referencedesign})
  \STATE Compute platoon change in control $\Delta \hat{U}_k^\star$ in \eqref{eq:platooncontrolminimisation} subject to state constraints \eqref{eq:qpconstraints}
  \STATE Compute latest control action $U_k$ in \eqref{eq:platoonchangeincontrolwithhuman}
  \STATE Step platoon dynamics \eqref{eq:platoonmodel}
  \ENDFOR
 \end{algorithmic}
 \end{algorithm}

%

\section{Numerical Study}
In this section, we provide guidance on the tuning of the cost function weights from \eqref{eq:costfuncwithheadway} and illustrate a numerical experiment.

\subsection{Cost Function Weights}

Increasing all of the penalties, $q_1$, $q_2$, $q_3$, and $q_4$, substantially can cause the optimization algorithm to become infeasible.
In the initial transient phase, it is necessary for the vehicles to deviate from the desired reference to enable convergence of both positions and velocities.
However, it is the initial transient phase where the impact is most prominent.

At minimum the position error penalties $q_1$ and $q_2$ are required to ensure that the platoon converges to the desired positions.
As motivated in \cite{Jovanovic2005IEEETransactionsonAutomaticControlillposednesscertain}, the absolute position error penalty $q_2$ must be present or convergence is a function of the platoon length.
As $q_1$ and $q_2$ affect the same position error states, it is suggested to tune these parameters together.
Increasing $q_2$ forces the vehicles to the position reference, with less regard to the relative distance, while increasing the relative position error penalty $q_1$ preferences the inter-vehicle distance over the absolute position reference.

It is possible to set the velocity ($q_3$) and acceleration ($q_4$) penalties to zero.
Convergence is natural following the position errors.
Increasing the velocity (acceleration) penalty forces the velocities (accelerations) closer to the desired reference, which can slow the convergence of the positions.

\subsection{Numerical Simulation}

We consider a numerical simulation of five ($M=5$) vehicles, with sampling period of $\Delta_t = 0.1$ [sec/sample].
We consider vehicle lengths as $l_i = 2.5$ [m] for all $i=\{1,5\}$, and the vehicle mechanical lags as $\tau_1 = 0.5$, $\tau_2 = 0.2$, $\tau_3 = 0.3$, $\tau_4 = 0.6$, and $\tau_5 = 0.4$ [sec].
The desired velocity is set as highway speed limit of $100$ [km/h] or $v_d = 27.78$ [m/s].
We consider that the drivers individually select their desired standstill distances as $r_1 = 6$, $r_2 = 6$, $r_3 = 5$, $r_4 = 8$, $r_5 = 7$ [m], and headway times as $h_k^{(1)} = 1$, $h_k^{(2)} = 0.4$, $h_k^{(3)} = 0.2$, $h_k^{(4)} = 0.3$, and $h_k^{(5)} = 1.4$ [sec].
At steady-state the inter-vehicle distances between the vehicles will be $\bar{d}_k^{(2)} = 19.61$, $\bar{d}_k^{(3)} = 13.06$, $\bar{d}_k^{(4)} = 18.83$, and $\bar{d}_k^{(5)} = 20.61$ [m].

At time $100$ [sec] the human driver of vehicle-$3$ performs an emergency brake to $0$ [m/s], then at $150$ [sec] increases speed to $11$ [m/s] until $250$ [sec] when the driver returns to platoon control.
Following the interruption to platoon automation from the driver of vehicle-3, the drivers decide to increase their individual inter-vehicle distances for additional safety and set their headway times to $h_k^{(2)} = 1.9$, $h_k^{(3)} = 1.7$, $h_k^{(4)} = 1.8$, and $h_k^{(5)} = 2.0$ [sec] at time $320$ [sec].
The new steady-state inter-vehicle distances will then be $\bar{d}_k^{(2)} = 61.28$, $\bar{d}_k^{(3)} = 54.72$, $\bar{d}_k^{(4)} = 60.50$, and $\bar{d}_k^{(5)} = 65.06$ [m].

We choose the acceleration constraints as $a_{\textrm{min}} = -6$ [m/s$^2$] and $a_{\textrm{max}} = 3$ [m/s$^2$], based on the performance of an average passenger vehicle and comfort of passengers.
We choose the velocity constraints as $v_{\textrm{min}} = 0$ [m/s] and $v_{\textrm{max}} = 27.8$ [m/s], based on the road speed limit.
Finally, the minimum inter-vehicle distance $d_{\textrm{min}} = 2$ [m] and maximum inter-vehicle distance $d_{\textrm{max}} = 70$ [m].
We consider a prediction horizon of $1.5$ [sec], or $15$ samples, with time to reach desired velocity of $40$ [sec] which equates to $k_m = 400$ [samples].

We choose the penalty on the
relative position errors or inter-vehicle distances as $q_1 = 1$,
absolute position error or error to the position reference as $q_2 = 1$,
velocity errors as $q_3 = 1$,
acceleration errors as $q_4 = 1$,
change in control as $R_\Delta = 2 I_M$, and
the terminal cost $P$ as the solution to the algebraic Riccatti equation.

\begin{figure}
\centering
\includegraphics[width=\columnwidth]{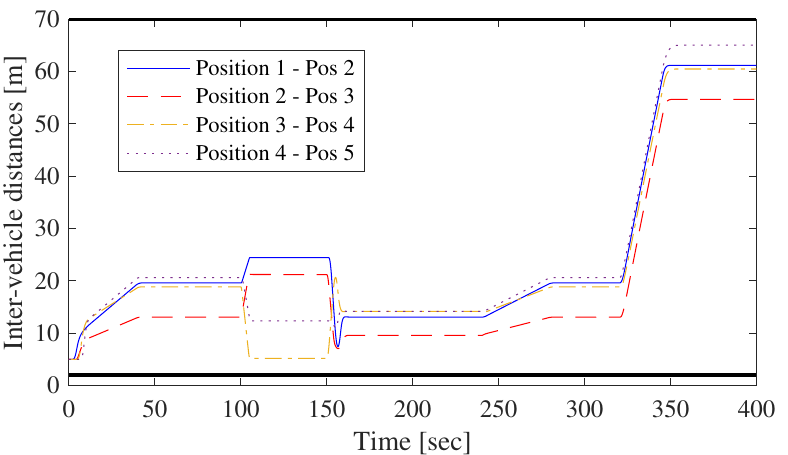}
\caption{Inter-vehicle Distance from simulation with five vehicles. The inter-vehicle distance smoothly converges to the desired values. The distances between each pair of neighboring vehicles are represented by colors (solid blue, dashed red, dash-dot yellow, dotted purple).}
\label{fig:MATLAB:intervehicledistance}
\end{figure}

We simulate our proposed control design using a standard convex optimizer\footnote{We use the MATLAB \texttt{mpcActiveSetSolver} from the \ac{MPC} toolbox.}
to solve the quadratic program \eqref{eq:quadraticprogramform} with constraints \eqref{eq:qpconstraints}, and human driver prediction \eqref{eq:humanmpc:costfunc} with \eqref{eq:humanmpc:constraints}.
The inter-vehicle distances are shown in Figure~\ref{fig:MATLAB:intervehicledistance}, with constraints as the solid horizontal black lines.
Our proposed controller converges the vehicles to the desired inter-vehicle distances by converging to the desired position reference and ensures the position constraints are maintained when vehicle-$3$ is controlled by a human driver.

\begin{figure}
\centering
\includegraphics[width=\columnwidth]{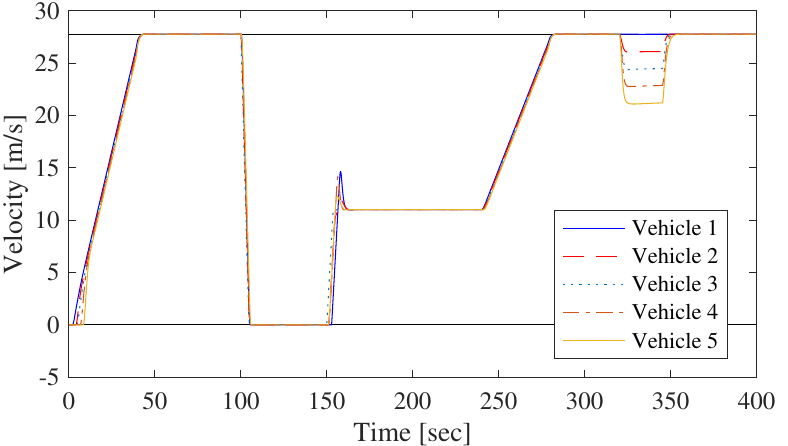}
\caption{Velocities of the five vehicles from the simulation. The vehicles follow the velocity reference and are represented by colors (solid blue, dashed red, dotted purple, dash-dot brown, solid gold).}
\label{fig:MATLAB:velocities}
\end{figure}

\begin{figure}
\centering
\includegraphics[width=\columnwidth]{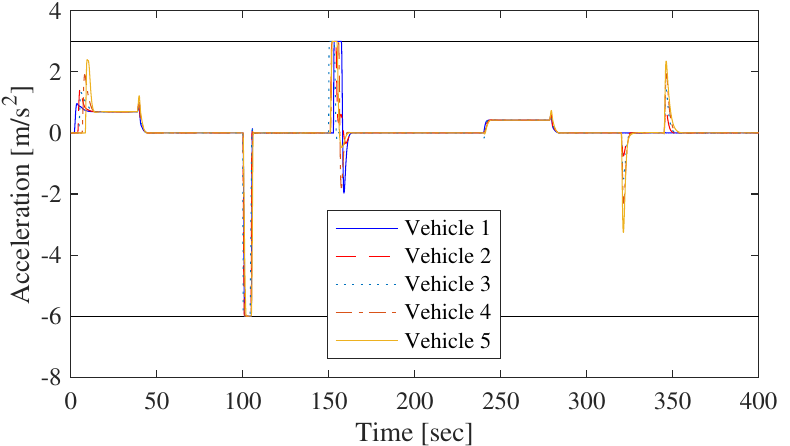}
\caption{Accelerations of the five vehicles from the simulation. The vehicles follow the acceleration reference and are represented by colors (solid blue, dashed red, dotted purple, dash-dot brown, solid gold).}
\label{fig:MATLAB:accelerations}
\end{figure}

Figure~\ref{fig:MATLAB:velocities} shows the velocities and Figure~\ref{fig:MATLAB:accelerations} shows the accelerations of the five vehicles in the platoon.
Our proposed controller smoothly converges the vehicles to the target velocity reference. We observe that the controller initially accelerates the latter vehicles in the platoon to converge to the target position references.

Our simulation results illustrate that our proposed control design successfully converges the controlled vehicles to the desired velocity and inter-vehicle distances.
The controller smoothly accelerates the vehicles to the desired position reference, minimizing the absolute and relative distance errors before converging the velocity to the reference.
The use of a constrained \ac{MPC} optimization approach ensures that safety margins on the inter-vehicle are maintained, the velocities are within the road speed limits, and the commanded accelerations are appropriate for the vehicle and comfortable for passengers.

Our simulation also illustrates that the use of a human driver model enables the centralized controller to operate in the presence of an unknown human driver.
Within one sample, our centralized controller reacts, reducing the speed of the remaining vehicles and ensuring that all vehicles in the platoon reach zero velocity before the inter-vehicle distance constraints are violated.
As the human controlled vehicle speeds up to a slower velocity than desired, the platoon then maintains the desired standstill distances with additional margin of the constant time headway while matching the lower velocity.
As the vehicle returns the platoon smoothly accelerates back to the desired velocity and returns to the full desired inter-vehicle distances.

\paragraph*{Remark}
A key contribution of our work is we include a human driver model within our centralized design \eqref{eq:mpc:predictionmodelfull}.
Our design avoids catastrophic safety incidents where the actions of the human driver, such as emergency braking, could result in an accident where a simpler platoon controller is unable to react to the human driver actions.
For example, during the emergency braking scenario a constant gain state feedback controller using only the current state in \eqref{eq:platoondynamics}, would command vehicles-4 and-5 to drive through vehicle-3, and give a constraint violating instruction to vehicle-3 when it re-joined the platoon.
An MPC designed for the platoon without the human driver model, see \eqref{eq:mpc:predictionmodel}, would become infeasible as the actions of the human driver in vehicle-3 would result in the state being constraint violating.

\paragraph*{Remark}
String stability has been well studied for linear systems and the so-called predecessor-follower topology \cite{Feng2019AnnualReviewsinControlStringstabilityvehicular,Besselink2016CyberPhysicalControlRoad,Wang2020SurveyCooperativeLongitudinal,Knorn2014AutomaticaPassivitybasedcontrol,Nunen2019StringStableModel}.
However, string stability for nonlinear systems and general topologies is very challenging, and remains an open problem \cite{Feng2019AnnualReviewsinControlStringstabilityvehicular}.
Due to the inclusion of constraints in the optimization problem \eqref{eq:qpconstraints}, the closed-loop dynamics resulting from our proposed controller are nonlinear.
Additionally, the centralized design with all vehicle states forms general design topology.
As our proposed centralized constrained controller is nonlinear with a general topology, showing string stability analytically is a non-trivial problem \cite{Feng2019AnnualReviewsinControlStringstabilityvehicular}.
String stability properties of our controller can be observed in simulation.

\begin{figure}
\centering
\includegraphics[width=\columnwidth]{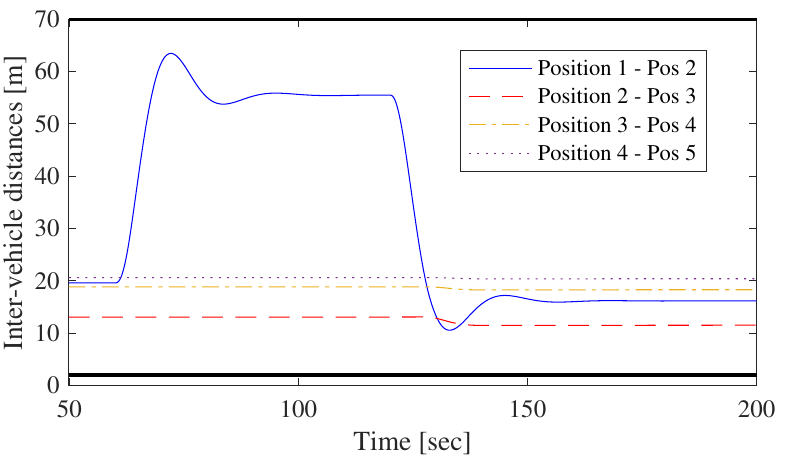}
\caption{Inter-vehicle Distance from simulation with five vehicles and disturbance to vehicle-1. The inter-vehicle distances start at steady state and diverge with application of the disturbance. The impact of the disturbance is reduced down the platoon indicating possible string stability. The distances between each pair of neighboring vehicles are represented by colors (solid blue, dashed red, dash-dot yellow, dotted purple).}
\label{fig:MATLAB:intervehicledistancewithdist}
\end{figure}

We illustrate in an extension to the above simulation example of our centralized controller for a platoon of five vehicles.
Consider the period just after the platoon reaches steady state, just after 40 [sec].
We apply a disturbance to vehicle-1 between 60 [sec] and 120 [sec].
Figure~\ref{fig:MATLAB:intervehicledistancewithdist} shows the inter-vehicle distance.
The impact on the inter-vehicle distance between vehicle-1 and-2 is clear (solid blue line).
However, the impact on the distance between vehicle-4 and-5 (dotted purple line) is very small.
We observe in simulation that the impact of the disturbance on the inter-vehicle distances reduces down the platoon, illustrating a key property of string stability.
Finally, when the disturbance is removed the platoon, the inter-vehicle distances return to the steady-state values.

\subsection{Comparison Simulation}
We provide a short comparison example by applying the continuous-time LQR controller of \cite{Jovanovic2005IEEETransactionsonAutomaticControlillposednesscertain} to our scenario above.
This controller is shown to be string stable and furthers the original centralized platoon control designs in \cite{Levine1966IEEETransactionsonAutomaticControloptimalerrorregulation} and \cite{Melzer1971AutomaticaOptimalregulationsystems}.
However, the control design does not take into account safety and actuation constraints, and for certain initial condition scenarios the control action may be larger than the possible actuation \cite{Jovanovic2004avoidingsaturationcontrol}.

In our comparison, we keep all scenario parameters outlined above the same, except as the design of \cite{Jovanovic2005IEEETransactionsonAutomaticControlillposednesscertain} does not incorporate individual inter-vehicle distances, we choose the largest initial steady-state distance considered above of $49.16$ [m] as the constant inter-vehicle distance.
Additionally, we note that the vehicle model, and hence controller, in \cite{Jovanovic2005IEEETransactionsonAutomaticControlillposednesscertain} does not include the acceleration, thus we consider the control action or desired velocity as the vehicle acceleration and saturate this control signal to the acceleration constraints above.
Finally, we note that there is no acceleration error penalty ($q_4$) or change in control ($R_\Delta$), while there is penalty on the control action which we choose as $r = 1$.

\begin{figure}
\centering
\includegraphics[width=\columnwidth]{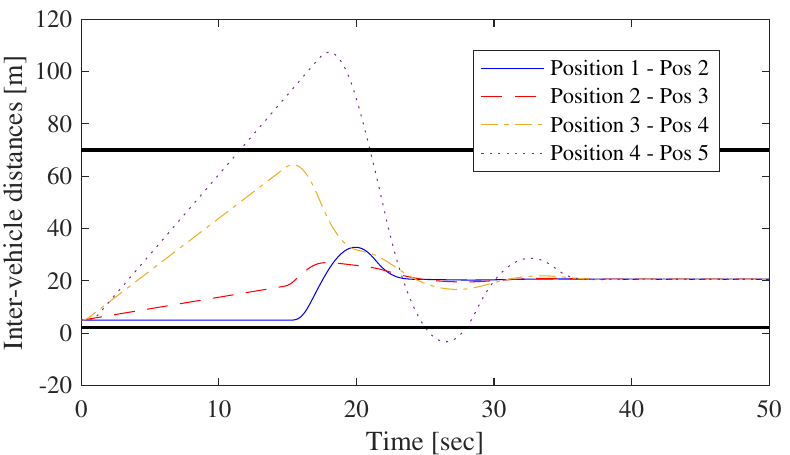}
\caption{Inter-vehicle Distance from the comparison simulation with five vehicles. Without inclusion of constraints in the control design, the inter-vehicle distance becomes quite large before vehicles-4 and-5 crash at 25.4 [sec]. The distances between each pair of neighboring vehicles are represented by colors (solid blue, dashed red, dash-dot yellow, dotted purple).}
\label{fig:comparison:intervehicledistance}
\end{figure}

\begin{figure}
\centering
\includegraphics[width=\columnwidth]{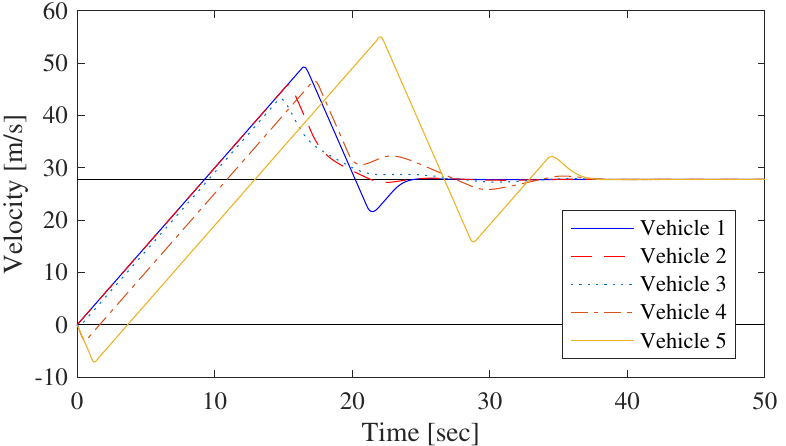}
\caption{Velocities of the five vehicles from the comparison simulation. The velocities become extremely large and are represented by colors (solid blue, dashed red, dotted purple, dash-dot brown, solid gold).}
\label{fig:comparison:velocities}
\end{figure}

Figures~\ref{fig:comparison:intervehicledistance} and~\ref{fig:comparison:velocities} show the inter-vehicle distances and velocities.
It is clear that the constraints, in black, are all violated very quickly.
We observe that the inter-vehicle distance between vehicles-4 and-5 becomes negative at 25.4 [sec], indicating that these vehicles have crashed.
Additionally, the velocities are absurd with vehicle-5 reaching velocities of -7.14 to 55.02 [m/s] (or -25.68 to 198.08 [km/h]).

In comparison, beyond incorporating human driving, a key contribution of our proposed controller is the inclusion of safety and actuation constraints.
Our control design converges the platoon of vehicles to the desired inter-vehicle distances and platoon velocity without violating the state and actuation constraints.
Additionally, we also include interruption from a human driver in our \ac{CACC} design.

%

\section{Conclusion}%
\label{sec:conclusion}

In this paper we propose a hybrid constrained \ac{MPC} algorithm to control a heterogeneous platoon of vehicles to a desired platoon velocity and inter-vehicle distance.
The finite horizon cost function of our centralized platoon controller is inspired from the infinite horizon cost function of \cite{Jovanovic2005IEEETransactionsonAutomaticControlillposednesscertain} with inclusion of headway times individual to each vehicle and able to be changed with time.
Through the use of constraints in the optimization, we ensure that the control actions result in safe vehicle behavior.
In our approach, we propose the use of a cost function to predict the control actions of a human driver that takes control of their vehicle, by assuming that the human driver will only obey at minimum, the legal velocity limits and the physical performance constraints of their vehicle.
We illustrate the performance of our control approach in a numerical study.
The centralized design allows the controller to utilize all possible vehicle state information.
Future work includes implementing distributed state estimation such that the centralized approach can be operated decentralized, with consideration of unreliable and delayed communication between vehicles.

%

%

\begin{IEEEbiography}[{\includegraphics[width=1in,height=1.25in,clip,keepaspectratio]{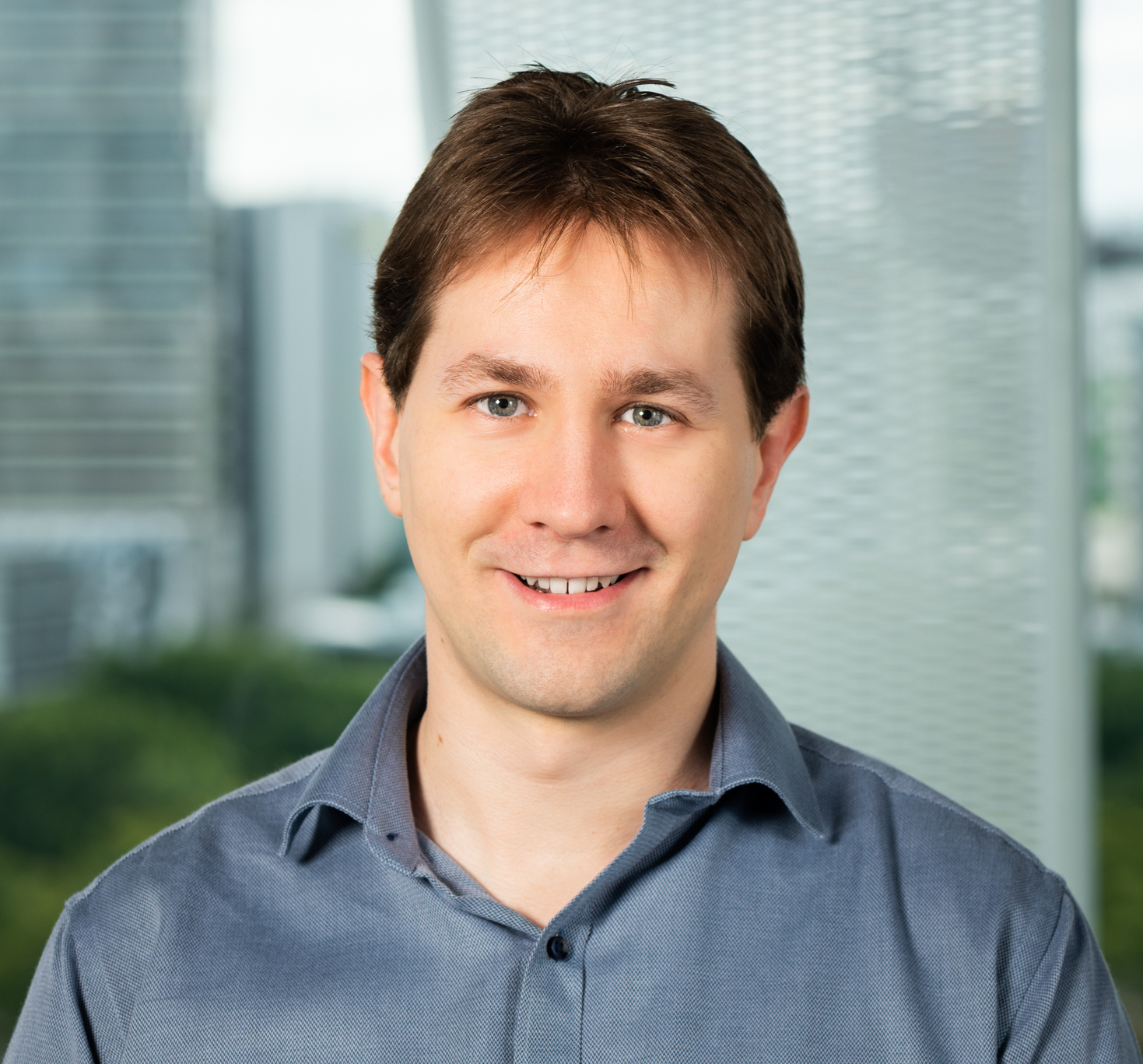}}]{Justin M. Kennedy}
(j12.kennedy@qut.edu.au)
is a Postdoctoral researcher with the School of Electrical Engineering and Robotics, Queensland University of Technology (QUT), Australia.
He received his B. Eng (Electrical)/B. Maths, and PhD degrees from QUT, in 2016 and 2022, respectively.
He recently completed his PhD into the estimation and control of marine craft in the presence of environmental disturbances.
Dr.\ Kennedy is a Member of IEEE, IEEE Control Systems Society (CSS), and
Society for Industrial and Applied Mathematics (SIAM).
His current research interest is in the application of mathematical and control system tools to solve network engineering problems.
\end{IEEEbiography}

\begin{IEEEbiography}[{\includegraphics[width=1in,height=1.25in,clip,keepaspectratio]{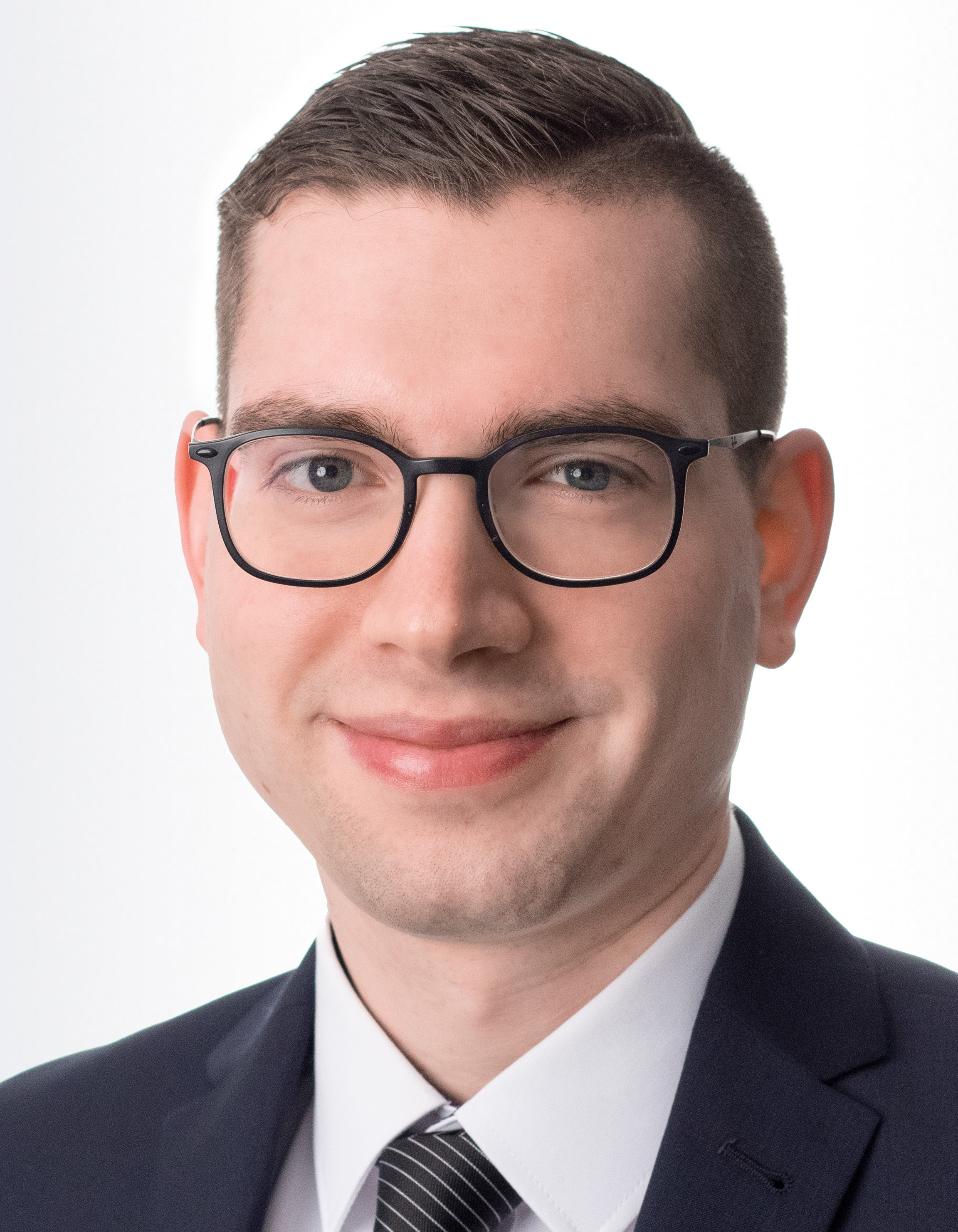}}]{Julian Heinovski}
(heinovski@ccs-labs.org)
is a PhD candidate and researcher at the Telecommunications Networks Group (TKN) at the School of Electrical Engineering and Computer Science, TU Berlin, Germany.
He received his B.Sc. and M.Sc. degrees from the Dept. of Computer Science, Paderborn University, Germany, in 2016 and 2018, respectively.
Julian is a Student Member of IEEE and ACM as well as a Member of IEEE Intelligent Transportation Systems Society (ITSS) and IEEE Vehicular Technology Society (VTS).
He serves as a reviewer of manuscripts in the field of vehicular networks and intelligent transportation systems.
His research interest is in cooperative driving as well as intelligent transportation systems, mainly focusing on platooning.
\end{IEEEbiography}

\begin{IEEEbiography}[{\includegraphics[width=1in,height=1.25in,clip,keepaspectratio]{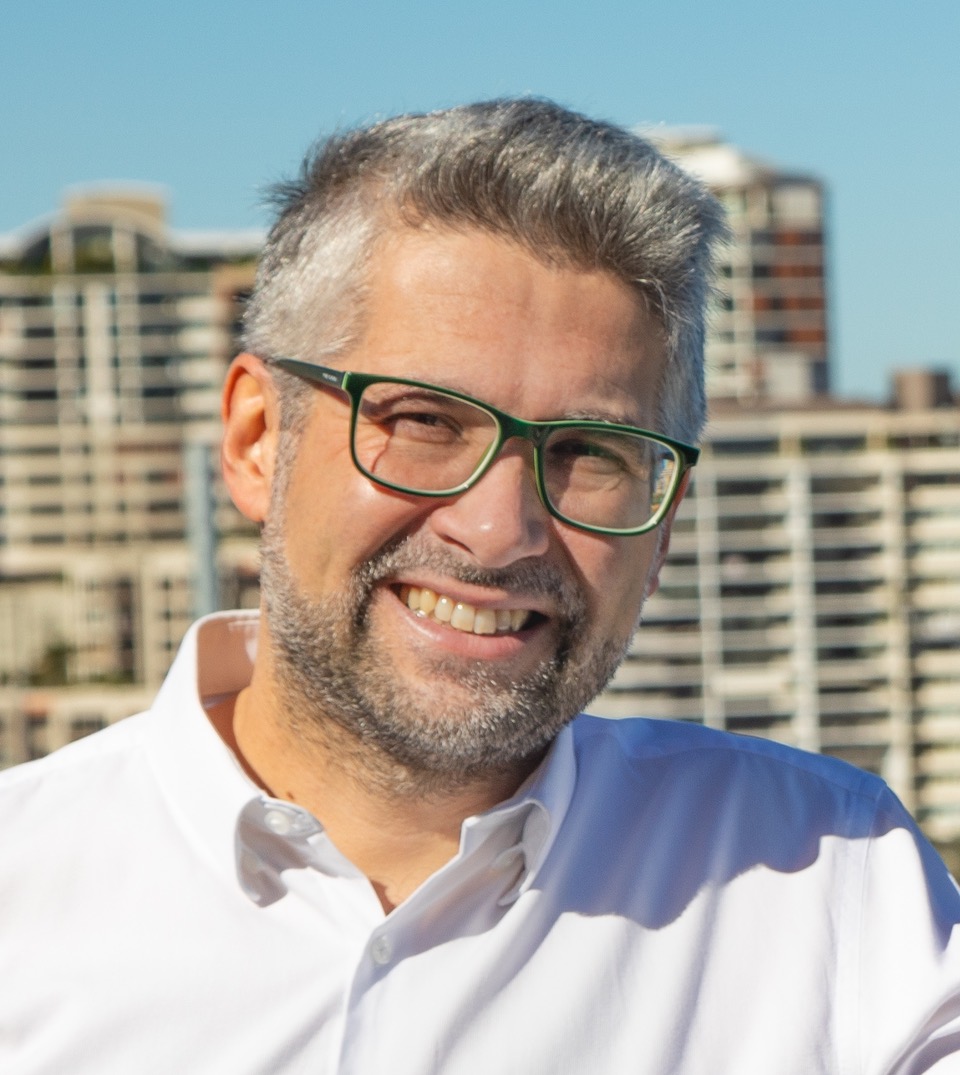}}]{Daniel E.\ Quevedo}
(dquevedo@ieee.org)
received Ingeniero Civil Electr\'onico
and M.Sc.\ degrees from   Universidad
T\'ecnica Federico Santa Mar\'{\i}a, Valpara\'{\i}so, Chile, in 2000, and in 2005   the Ph.D.\ degree from the University of Newcastle, Australia.
He is Professor of Cyberphysical Systems
at the School of Electrical Engineering and Robotics, Queensland University of Technology (QUT), 
in Australia.
Before joining QUT, he established and led  the Chair in Automatic
Control at Paderborn University, Germany.
He is co-recipient of the  2018 IEEE Transactions on Automatic Control George S.\ Axelby Outstanding Paper Award.
\par Prof.\ Quevedo currently serves as Associate Editor for   \emph{IEEE Control Systems}.
From 2015--2018 he was Chair of the IEEE Control Systems Society \emph{Technical Committee on
  Networks \& Communication Systems}.
His research interests are in networked control systems and cyberphysical systems security.
\end{IEEEbiography}

\begin{IEEEbiography}[{\includegraphics[width=1in,height=1.25in,clip,keepaspectratio]{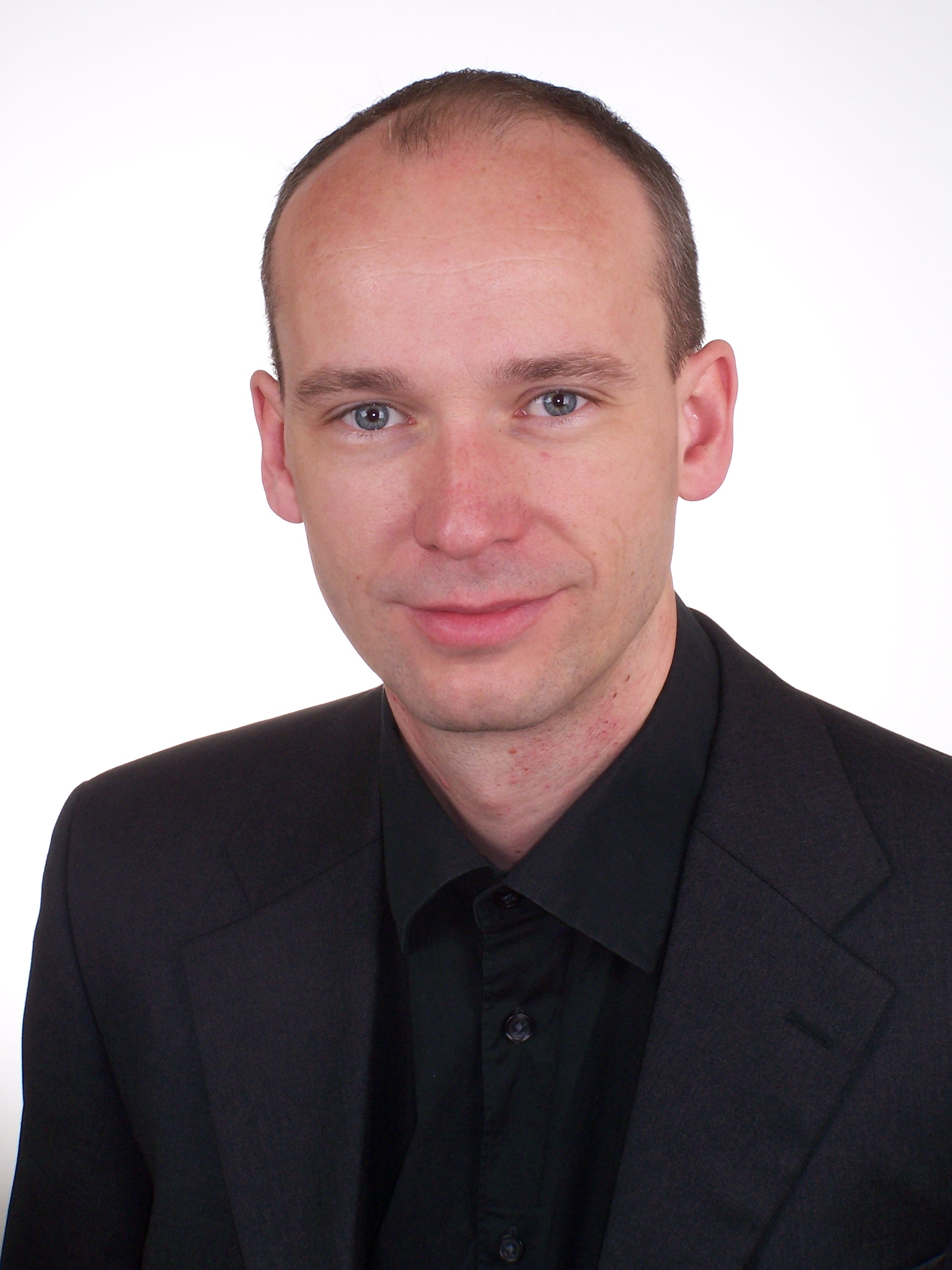}}]{Falko Dressler}
(dressler@ccs-labs.org)
is full professor and Chair for Telecommunication Networks at the School of Electrical Engineering and Computer Science, TU Berlin. He received his M.Sc. and Ph.D. degrees from the Dept. of Computer Science, University of Erlangen in 1998 and 2003, respectively.
Dr. Dressler has been associate editor-in-chief for IEEE Trans. on Mobile Computing and Elsevier Computer Communications as well as an editor for journals such as IEEE/ACM Trans. on Networking, IEEE Trans. on Network Science and Engineering, Elsevier Ad Hoc Networks, and Elsevier Nano Communication Networks. He has been chairing conferences such as IEEE INFOCOM, ACM MobiSys, ACM MobiHoc, IEEE VNC, IEEE GLOBECOM. He authored the textbooks Self-Organization in Sensor and Actor Networks published by Wiley \& Sons and Vehicular Networking published by Cambridge University Press. He has been an IEEE Distinguished Lecturer as well as an ACM Distinguished Speaker.
Dr. Dressler is an IEEE Fellow as well as an ACM Distinguished Member. He is a member of the German National Academy of Science and Engineering (acatech). He has been serving on the IEEE COMSOC Conference Council and the ACM SIGMOBILE Executive Committee. His research objectives include adaptive wireless networking (sub-6GHz, mmWave, visible light, molecular communication) and wireless-based sensing with applications in ad hoc and sensor networks, the Internet of Things, and Cyber-Physical Systems.
\end{IEEEbiography}

%

\section{Appendix}

\subsection{Single Vehicle Dynamics}
\label{sec:app:singledynamics}

The continuous-time dynamics and control input matrices of the dynamics \eqref{eq:singlecontinuousdynamics} are
\begin{equation*}
	A_c^{(i)} \triangleq \begin{bmatrix} 0 & 1 & 0 \\ 0 & 0 & 1 \\ 0 & 0 & -\frac{1}{\tau_i} \end{bmatrix}, \quad \textrm{and} \quad
	B_c^{(i)} \triangleq \begin{bmatrix} 0 \\ 0 \\ \frac{1}{\tau_i} \end{bmatrix} .
\end{equation*}

Following \cite{AAstrom2011ComputerControlledSystems}, a continuous-time system can be discretized with sampling interval $\Delta_t$ [s] using
\begin{align*}
	A^{(i)} = \exp(& A_c^{(i)} \Delta_t) ~ \textrm{and} ~ B^{(i)} = \int_0^{\Delta_t} \exp(A_c^{(i)} m) dm ~ B_c^{(i)} \\
\textrm{to give }
	A^{(i)} &= \begin{bmatrix} 1 & \Delta_t & \tau_i \left(\Delta_t - \tau_i \left(1 - \exp\left(-\frac{\Delta_t}{\tau_i}\right)\right)\right) \\ 0 & 1 & \tau_i \left(1 - \exp\left(-\frac{\Delta_t}{\tau_i}\right)\right) \\ 0 & 0 & \exp\left(-\frac{\Delta_t}{\tau_i}\right) \end{bmatrix} \\
\textrm{and  }
	B^{(i)} &= \begin{bmatrix} -\tau_i \left(\Delta_t - \tau_i \left(1 - \exp\left(\frac{-\Delta_t}{\tau_i}\right)\right)\right) + \frac{\Delta_t^2}{2} \\ \Delta_t - \tau_i \left(1 - \exp\left(\frac{-\Delta_t}{\tau_i}\right)\right) \\ 1 - \exp\left(\frac{-\Delta_t}{\tau_i}\right) \end{bmatrix} .
\end{align*}

\subsection{Platoon Dynamics}
\label{sec:app:platoondynamics}
The block diagonal dynamics matrices of \eqref{eq:platoondynamics} are defined as
\begin{equation*}
    A_M = \begin{bmatrix}
        I_M & \Delta_t I_M & A_M^{(1,3)} \\
        0 & I_M & A_M^{(2,3)} \\
        0 & 0 & A_M^{(3,3)}
    \end{bmatrix}
    \quad \textrm{and} \quad
    B_M = \begin{bmatrix}
    B_M^{(1,1)} \\ B_M^{(2,1)} \\ B_M^{(3,1)}
    \end{bmatrix}
\end{equation*}
where $I_M$ the identity matrix of size $M \times M$, $0$ is a matrix of zeros of appropriate size, and
\begin{align*}
    A_M^{(1,3)} &= \textrm{diag}\left[\tau_1 \left(\Delta_t - \tau_1 \left(1 - \exp\left(-\frac{\Delta_t}{\tau_1}\right)\right)\right), \dots, \right. \\ &\qquad\qquad\left. \tau_M \left(\Delta_t - \tau_M \left(1 - \exp\left(-\frac{\Delta_t}{\tau_M}\right)\right)\right) \right] \\
    A_M^{(2,3)} &= \textrm{diag} \left[
    \tau_1 \left(1 - \exp\left(-\frac{\Delta_t}{\tau_1}\right)\right), \dots, \right. \\ &\qquad\qquad\left. \tau_M \left(1 - \exp\left(-\frac{\Delta_t}{\tau_M}\right)\right) \right] \\
    A_M^{(3,3)} &= \textrm{diag} \left[ \exp\left(-\frac{\Delta_t}{\tau_1}\right), \dots, \exp\left(-\frac{\Delta_t}{\tau_M}\right) \right] \\
    B_M^{(1,1)} &= \textrm{diag} \left[ -\tau_1 \left(\Delta_t - \tau_1 \left(1 - \exp\left(\frac{-\Delta_t}{\tau_1}\right)\right)\right) + \frac{\Delta_t^2}{2}, \right. \\ &\quad\left. \dots, -\tau_M \left(\Delta_t - \tau_M \left(1 - \exp\left(\frac{-\Delta_t}{\tau_M}\right)\right)\right) + \frac{\Delta_t^2}{2} \right] \\
    B_M^{(2,1)} &= \textrm{diag} \left[ \Delta_t - \tau_1 \left(1 - \exp\left(\frac{-\Delta_t}{\tau_1}\right)\right), \dots, \right. \\ &\qquad\qquad\left. \Delta_t - \tau_M \left(1 - \exp\left(\frac{-\Delta_t}{\tau_M}\right)\right) \right] \\
    B_M^{(3,1)} &= \textrm{diag} \left[ 1 - \exp\left(\frac{-\Delta_t}{\tau_1}\right), \dots, 1 - \exp\left(\frac{-\Delta_t}{\tau_M}\right) \right]
\end{align*}
In the case of a homogeneous platoon, $\tau_i = \tau$, then $A^{(i)} = A$ and $B^{(i)} = B$ and the platoon dynamics can be conveniently computed $A_M = A \otimes I_M$ and $B_M = B \otimes I_M$, where $\otimes$ is the Kronecker operator.

The matrices of the prediction model of the platoon dynamics \eqref{eq:mpc:predictionmodel} are
\begin{align*}
	&\Phi = \begin{bmatrix} A_M \\ \vdots \\ A_M^N \end{bmatrix}, \quad
	\lambda = \begin{bmatrix} A_M^0 B_M \\ \vdots \\ (A_M^{N-1} + \dots + A_M^0) B_M \end{bmatrix}, \\
	&\textrm{and} \quad \Gamma = \begin{bmatrix}
		B_M & \cdots & 0 \\
		\vdots & \ddots & \vdots \\
		(A_M^{N-1} + \dots + A_M^0)B_M & \cdots & B_M
	\end{bmatrix}
\end{align*}
and for the human controlled vehicle \eqref{eq:humandriverdynamics} are
\begin{align*}
	&\bar{\Phi} = \begin{bmatrix} A^{(\ell)} \\ \vdots \\ (A^{(\ell)})^N \end{bmatrix} \quad
	\bar{\lambda} = \begin{bmatrix} (A^{(\ell)})^0 B^{(\ell)} \\ \vdots \\ ((A^{(\ell)})^{N-1} + \dots + (A^{(\ell)})^0) B^{(\ell)} \end{bmatrix}, \\
    &\textrm{and} \quad
    \bar{\Gamma} = \begin{bmatrix}
		B^{(\ell)} & \cdots & 0 \\
		\vdots & \ddots & \vdots \\
		((A^{(\ell)})^{N-1} + \dots + (A^{(\ell)})^0)B^{(\ell)} & \cdots & B^{(\ell)}
	\end{bmatrix} .
\end{align*}

\subsection{Cost Function Expansion}
\label{sec:app:costfunc}
In the below, we show the expansion of sums in the cost function from \eqref{eq:costfuncwithheadway} to the matrix version \eqref{eq:condensedcostfunc}.
Consider \eqref{eq:costfuncwithheadway}
\begin{align*}
	J &= \sum_{j=0}^{N-1} \left[ \sum_{i=1}^{M+1} q_1 \left(\hat{\eta}_{k+j|k}^{(i)}\right)^2 + \sum_{i=1}^M \left(q_2 \left(\hat{\xi}_{k+j|k}^{(i)}\right)^2 \right.\right.\nonumber\\&\quad\left.\left.+ q_3 \left(\hat{\zeta}_{k+j|k}^{(i)}\right)^2 + q_4 \left(\hat{\psi}_{k+j|k}^{(i)}\right)^2 + r \left(\Delta u_{k+j|k}^{(i)}\right)^2 \right) \right] \\ &\quad+ (\hat{X}_{k+N|k} - X_{k+N}^\star)^\mathsf{T} P_{k+N} (\hat{X}_{k+N|k} - X_{k+N}^\star)
\end{align*}

The first sum in \eqref{eq:costfuncwithheadway} is of the relative position errors $\hat{\eta}_\kappa$ noting that we substitute $\kappa$ in place of $k+j|k$.
We start by showing this in terms of the absolute position errors and velocity errors
\begin{align*}
    &\sum_{i=1}^{M+1} q_1 \left(\hat{\eta}_\kappa^{(i)}\right)^2
    = q_1 \sum_{i=1}^{M+1} \left(\hat{\xi}_\kappa^{(i)} - \hat{\xi}_\kappa^{(i-1)} + h_\kappa^{(i)} \hat{\zeta}_\kappa^{(i)} \right)^2 \\
    = &q_1 \sum_{i=1}^{M+1} \left[ \left(\hat{\xi}_\kappa^{(i)}\right)^2 + \left(\hat{\xi}_\kappa^{(i-1)}\right)^2 + \left(h_\kappa^{(i)}\right)^2 \left(\hat{\zeta}_\kappa^{(i)} \right)^2 \right. \\ &\left.\quad- 2 \hat{\xi}_\kappa^{(i)} \hat{\xi}_\kappa^{(i-1)} + 2 h_\kappa^{(i)} \hat{\xi}_\kappa^{(i)} \hat{\zeta}_\kappa^{(i)} - 2 h_\kappa^{(i)} \hat{\xi}_\kappa^{(i-1)} \hat{\zeta}_\kappa^{(i)} \right] \\
    = &q_1 \left[\left(\hat{\xi}_\kappa^{(1)}\right)^2 + \left(\hat{\xi}_\kappa^{(0)}\right)^2 + \left(h_\kappa^{(1)}\right)^2 \left(\hat{\zeta}_\kappa^{(1)} \right)^2 \right. \\ &\quad- 2 \hat{\xi}_\kappa^{(1)} \hat{\xi}_\kappa^{(0)} + 2 h_\kappa^{(1)} \hat{\xi}_\kappa^{(1)} \hat{\zeta}_\kappa^{(1)} - 2 h_\kappa^{(1)} \hat{\xi}_\kappa^{(0)} \hat{\zeta}_\kappa^{(1)}  \\
    &+\left(\hat{\xi}_\kappa^{(2)}\right)^2 + \left(\hat{\xi}_\kappa^{(1)}\right)^2 + \left(h_\kappa^{(2)}\right)^2 \left(\hat{\zeta}_\kappa^{(2)} \right)^2 \\ &\quad- 2 \hat{\xi}_\kappa^{(2)} \hat{\xi}_\kappa^{(1)} + 2 h_\kappa^{(2)} \hat{\xi}_\kappa^{(2)} \hat{\zeta}_\kappa^{(2)} - 2 h_\kappa^{(2)} \hat{\xi}_\kappa^{(1)} \hat{\zeta}_\kappa^{(2)} \\
    &+ \dots + \left(\hat{\xi}_\kappa^{(M)}\right)^2 + \left(\hat{\xi}_\kappa^{(M-1)}\right)^2 + \left(h_\kappa^{(M)}\right)^2 \left(\hat{\zeta}_\kappa^{(M)} \right)^2 \\ &\quad- 2 \hat{\xi}_\kappa^{(M)} \hat{\xi}_\kappa^{(M-1)} + 2 h_\kappa^{(M)} \hat{\xi}_\kappa^{(M)} \hat{\zeta}_\kappa^{(M)} - 2 h_\kappa^{(M)} \hat{\xi}_\kappa^{(M-1)} \hat{\zeta}_\kappa^{(M)} \\
    &+ \left(\hat{\xi}_\kappa^{(M+1)}\right)^2 + \left(\hat{\xi}_\kappa^{(M)}\right)^2 + \left(h_\kappa^{(M+1)}\right)^2 \left(\hat{\zeta}_\kappa^{(M+1)} \right)^2 \\ &\quad - 2 \hat{\xi}_\kappa^{(M+1)} \hat{\xi}_\kappa^{(M)} + 2 h_\kappa^{(M+1)} \hat{\xi}_\kappa^{(M+1)} \hat{\zeta}_\kappa^{(M+1)} \\ &\quad\left.- 2 h_\kappa^{(M+1)} \hat{\xi}_\kappa^{(M)} \hat{\zeta}_\kappa^{(M+1)} \right]
\end{align*}
Recall that the virtual lead and tail vehicles perfectly follow the reference such that $\hat{\xi}_\kappa^{(0)} = \hat{\xi}_\kappa^{(M+1)} = 0,$ and $\hat{\zeta}_\kappa^{(0)} = \hat{\zeta}_\kappa^{(M+1)} = 0,$ then
\begin{align*}
    = &q_1 \left[ 2 \left(\hat{\xi}_\kappa^{(1)}\right)^2 + \dots + \left(\hat{\xi}_\kappa^{(M)}\right)^2 \right.\\ &\quad- 2 \hat{\xi}_\kappa^{(2)} \hat{\xi}_\kappa^{(1)} - \dots - 2 \hat{\xi}_\kappa^{(M)} \hat{\xi}_\kappa^{(M-1)} \\
    & + \left(h_\kappa^{(1)}\right)^2 \left(\hat{\zeta}_\kappa^{(1)} \right)^2 + \dots + \left(h_\kappa^{(M)}\right)^2 \left(\hat{\zeta}_\kappa^{(M)} \right)^2 \\
    &+ 2 h_\kappa^{(1)} \hat{\xi}_\kappa^{(1)} \hat{\zeta}_\kappa^{(1)} + \dots + 2 h_\kappa^{(M)} \hat{\xi}_\kappa^{(M)} \hat{\zeta}_\kappa^{(M)} \\ &\quad\left.- 2 h_\kappa^{(2)} \hat{\xi}_\kappa^{(1)} \hat{\zeta}_\kappa^{(2)} - \dots - 2 h_\kappa^{(M)} \hat{\xi}_\kappa^{(M-1)} \hat{\zeta}_\kappa^{(M)} \right]
\end{align*}

This can be written in matrix notation as
\begin{align*}
    &q_1 \left(\hat{\xi}_\kappa^\mathsf{T} T_M \hat{\xi}_\kappa + \hat{\zeta}_\kappa^\mathsf{T} H_\kappa \hat{\zeta}_\kappa + \hat{\xi}^\mathsf{T} T_\kappa \hat{\zeta}_\kappa + \hat{\zeta}_\kappa^\mathsf{T} T_\kappa^\mathsf{T} \hat{\xi}_\kappa  \right) \\
    = &q_1 \begin{bmatrix} \hat{\xi}_\kappa \\ \hat{\zeta}_\kappa \end{bmatrix}^\mathsf{T}
    \begin{bmatrix} T_M & T_\kappa \\ T_\kappa^\mathsf{T} & H_\kappa \end{bmatrix}
    \begin{bmatrix} \hat{\xi}_\kappa \\ \hat{\zeta}_\kappa \end{bmatrix}
\end{align*}
where $T_M$ is a symmetric Toeplitz matrix of size $M \times M$ with the first row of the form $[2, -1, 0, \dots, 0]$, and $T_\kappa$ and $H_\kappa$ are $M \times M$ matrices where
\begin{align*}
    &T_\kappa = \begin{bmatrix}
        h_\kappa^{(1)} & -h_\kappa^{(2)} & 0 & \dots & 0 & 0 \\
        0 & h_\kappa^{(2)} & -h_\kappa^{(3)} & \dots & 0 & 0 \\
        \vdots &  & \ddots & \ddots &  & \vdots \\
        \vdots &  &  & \ddots & \ddots & \vdots \\
        0 & 0 & 0 & \dots & h_\kappa^{(M-1)} & -h_\kappa^{(M)} \\
        0 & 0 & 0 & \dots & 0 & h_\kappa^{(M)}
    \end{bmatrix} \\ &\textrm{and} \quad
    H_\kappa = 
    \textrm{diag} ~ \begin{bmatrix}
    \left(h_\kappa^{(1)}\right)^2 , \dots , \left(h_\kappa^{(M)}\right)^2
    \end{bmatrix}.
\end{align*}

Returning to the full sum, we can see that the other terms can also be written in matrix notation
\begin{align*}
    J &= \sum_{j=0}^{N - 1} \left[ q_1 \begin{bmatrix} \hat{\xi}_{k+j|k} \\ \hat{\zeta}_{k+j|k} \end{bmatrix}^\mathsf{T}
    \begin{bmatrix} T_M & T_{k+j} \\ T_{k+j}^\mathsf{T} & H_{k+j} \end{bmatrix}
    \begin{bmatrix} \hat{\xi}_{k+j|k} \\ \hat{\zeta}_{k+j|k} \end{bmatrix} \right.\\&\qquad\left.+ q_2 \hat{\xi}_{k+j|k}^\mathsf{T} I_M \hat{\xi}_{k+j|k} + q_3 \hat{\zeta}_{k+j|k}^\mathsf{T} I_M \hat{\zeta}_{k+j|k} \right.\\&\qquad\left.+ q_4 \hat{\psi}_{k+j|k}^\mathsf{T} I_m \hat{\psi}_{k+j|k} + r \left(\Delta U_{k+j|k}\right)^\mathsf{T} I_M \Delta U_{k+j|k} \right] \\
    &\quad+ (\hat{X}_{k+N|k} - X_{k+N}^\star)^\mathsf{T} P_{k+N} (\hat{X}_{k+N|k} - X_{k+N}^\star) \\
    &= \sum_{j=0}^{N - 1} \left[ \begin{bmatrix} \hat{\xi}_{k+j|k} \\ \hat{\zeta}_{k+j|k} \\ \hat{\psi}_{k+j|k} \end{bmatrix}^\mathsf{T} Q_{k+j} 
    \begin{bmatrix} \hat{\xi}_{k+j|k} \\ \hat{\zeta}_{k+j|k} \\ \hat{\psi}_{k+j|k} \end{bmatrix} \right.\\&\quad\left.+ \left(\Delta U_{k+j|k}\right)^\mathsf{T} R_\Delta \Delta U_{k+j|k} \right] \\&\quad+ (\hat{X}_{k+N|k} - X_{k+N}^\star)^\mathsf{T} P_{k+N} (\hat{X}_{k+N|k} - X_{k+N}^\star)
\end{align*}
where from \eqref{eq:beststatepenalty} of $R_\Delta = r I_M$ and
\begin{equation*}
	Q_\kappa = \begin{bmatrix} q_1 T_M + q_2 I_M & q_1 T_\kappa & 0 \\ q_1 T_\kappa^\mathsf{T} & q_1 H_\kappa + q_3 I_M & 0 \\ 0 & 0 & q_4 I_M \end{bmatrix}
\end{equation*}
then with further simplification of state and reference we find the cost function in reduced form \eqref{eq:condensedcostfunc}.

\end{document}